\documentclass[12pt, oneside]{article}

\usepackage{amsmath, amssymb, amsthm, amsbsy}
\usepackage[mathscr]{euscript} 
\usepackage{stmaryrd} %
\usepackage{bbm}
\usepackage[dvipsnames]{xcolor}
\usepackage{graphicx}
\usepackage{epigraph}
\usepackage{multicol} %
\usepackage{caption} %
\usepackage{subcaption} %
\usepackage[left=1.2in, right=1.2in, top=1.0in, bottom=0.8in, includehead, includefoot]{geometry}
\usepackage{setspace}
\usepackage{booktabs}
\usepackage{array}
\usepackage{tablefootnote} %
\usepackage{multirow}
\usepackage{float}
\usepackage[section]{placeins}

\usepackage{fontspec} 
\usepackage[xcharter]{newtxmath}
\usepackage{imakeidx}

\usepackage{authorindex}

\usepackage[colorlinks=true, citecolor=Blue, linkcolor=blue]{hyperref}
\newcommand\myshade{85}
\colorlet{mylinkcolor}{violet}
\colorlet{mycitecolor}{PineGreen}
\colorlet{myurlcolor}{Aquamarine}
\hypersetup{
  linkcolor  = mylinkcolor!\myshade!black,
  citecolor  = mycitecolor!\myshade!black,
  urlcolor   = myurlcolor!\myshade!black,
  colorlinks = true,
}

\usepackage[lastexercise, answerdelayed]{exercise}

\usepackage{chngcntr}
\counterwithin{Exercise}{section}
\counterwithin{Answer}{section}

\usepackage{natbib}

\usepackage{enumitem}
\setlist[enumerate]{itemsep=2pt,topsep=3pt}
\setlist[itemize]{itemsep=2pt,topsep=3pt}
\setlist[enumerate,1]{label={\upshape (\roman*)}}

\usepackage[linesnumbered, ruled]{algorithm2e}
\SetAlCapSkip{1.2em}

\usepackage{tikz}
\usepackage{tikz-cd}
\usetikzlibrary{positioning}
\usetikzlibrary{arrows}
\usetikzlibrary{calc}
\usetikzlibrary{intersections}
\usetikzlibrary{matrix}
\usetikzlibrary{decorations}
\usepackage{pgf}
\usepackage{pgfplots}
\pgfplotsset{compat=1.16} %
\usetikzlibrary{shapes, fit}
\usetikzlibrary{arrows.meta} %
\usetikzlibrary{decorations.pathreplacing}  %
\usepackage{pgfgantt}

\definecolor{codebg}{rgb}{0.95, 0.93, 0.93}
\definecolor{pale}{RGB}{235, 235, 235}
\definecolor{pale2}{RGB}{255,255,255}
\definecolor{aquamarine}{RGB}{69,139,116}

\newcommand{\EE}{\mathbbm E}

\newcommand{\RR}{\mathbbm R}

\newcommand{\PP}{\mathbbm P}

\renewcommand{\leq}{\leqslant}
\renewcommand{\geq}{\geqslant}

\renewcommand{\phi}{\varphi}
\renewcommand{\epsilon}{\varepsilon}

\newcommand{\cf}[1]{ \lstinline|#1| }

\theoremstyle{plain}

\newtheorem{theorem}{Theorem}[section]
\newtheorem{corollary}[theorem]{Corollary}
\newtheorem{lemma}[theorem]{Lemma}
\newtheorem{proposition}[theorem]{Proposition}

\theoremstyle{definition}
\newtheorem{definition}{Definition}[section]

\newtheorem{remark}{Remark}[section]
\newtheorem{assumption}{Assumption}[section]
\newtheorem{condition}[assumption]{Condition}

\newcommand{\HildaWaveTwentyTwoAlphaConsumption}{3.168}
\newcommand{\HildaWaveTwentyTwoAlphaResource}{2.119}
\newcommand{\HildaWaveTwentyTwoGap}{1.050}

\newcommand{\HildaWaveTwentyTwoSEGap}{0.199}

\newcommand{\HfcsDeAlphaResource}{1.603}
\newcommand{\HfcsDeGap}{2.482}

\newcommand{\HfcsDeSEGap}{0.276}

\newcommand{\HfcsItGap}{1.768}

\newcommand{\HfcsItSEGap}{0.223}

\newcommand{\ChipsTwentyThirteenGap}{1.012}

\newcommand{\ChipsTwentyThirteenSEGap}{0.108}

\newcommand{\ChipsTwentyEighteenGap}{1.195}

\newcommand{\ChipsTwentyEighteenSEGap}{0.065}

\newcommand{\ChThreeDeRho}{0.392}

\newcommand{\ChThreeNlRho}{0.722}

\title{Consumption and Wealth Tails in Heterogeneous-Agent Models:\\ The Pareto Trap}

\author{Shu Hu\thanks{I thank John Stachurski for supervision, detailed comments, and guidance. My special thanks go to Thomas J. Sargent, who sparked the idea for this paper during an afternoon walk at the ANU in January 2023. I am also grateful to Timothy Kam, Fedor Iskhakov, Christopher Carroll, Serguei Maliar, Chung Tran, Juergen Meinecke, Elena Capatina, Jingni Yang, Humphrey Yang, Charlie Xia, James Graham, Xiaodong Fan, Aniket Baksy, and audiences at the ANU RSE Annual PhD in Economics Workshop in 2023 and 2024, at the 39th Australian PhD Conference at UNSW, and at ESAM 2026 for helpful questions and comments. Any remaining errors are my own.}\\
\small Australian National University\\
\small \href{mailto:shu.hu@anu.edu.au}{shu.hu@anu.edu.au}}

\date{November 2026}

\begin{document}

\maketitle

\begin{abstract}
Household wealth is more concentrated at the top than household consumption. In weighted surveys for Australia, Germany, Italy, and the Netherlands, the fitted consumption tail index exceeds the wealth index, and three of four index gaps survive a prespecified multiple-comparison adjustment. Benchmark heterogeneous-agent models cannot reproduce this ranking: whenever optimal consumption remains above a fixed positive fraction of wealth, a Pareto-type wealth tail keeps the consumption tail at least as heavy, a \emph{Pareto trap}. The restriction extends to a finitely lived overlapping-generations economy. Calibrated to each eligible wealth index, the benchmark's untargeted consumption predictions miss all four point estimates.

\end{abstract}

\section{Introduction}
\label{sec:intro}

Household wealth is substantially more concentrated at the top than household
consumption.  Write $\alpha_C$ and $\alpha_W$ for the population Pareto tail
indices and $\hat\alpha_C$ and $\hat\alpha_W$ for their estimates.  A smaller
index means a heavier tail.  Existing U.S.\ evidence
places the wealth index near $1.5$ and the consumption index around $3$
\citep{klass2007forbes,vermeulen2016estimating,
gaillard2023consumption}.  The estimated ranking is therefore
$\hat\alpha_C>\hat\alpha_W$.  This paper asks what that ranking implies for
heterogeneous-agent models calibrated to the wealth tail.

The key observation is an overidentifying restriction.  Suppose a model keeps
consumption above a positive fraction of a household resource variable at high
resource levels.  Then households in the resource tail necessarily populate the
consumption tail, so consumption cannot have a strictly thinner Pareto tail.
Once the model is disciplined to match $\alpha_W$, it cannot choose an
arbitrary $\alpha_C$ while this exposure persists.  Matching the wealth tail
is therefore not sufficient model validation.  It creates a second, untargeted
implication for consumption.

The paper makes three contributions.  First, it estimates both tail indices
in three household survey systems spanning eight countries (HILDA, the
euro-area HFCS, and CHIP), complementing the U.S.\ evidence in
\citet{gaillard2023consumption}.  The fitted consumption index exceeds the
fitted wealth index in every weighted-survey comparison, four index pairs
pass the prespecified eligibility checks, and the difference survives the
prespecified Holm adjustment for Australia, Germany, and Italy
(Section~\ref{sec:empirics}).  The two Chinese nonhousing-resource
comparisons are reported over the observed range only.  These estimates and
their eligibility rules are fixed before any model comparison.

Second, Theorem~\ref{thm:3} derives the tail restriction from one property of
the consumption policy.  If $C\geq aX$ once resources $X$ are sufficiently
large, the event $\{X>c/a\}$ is contained in $\{C>c\}$.  Thus a Pareto
resource tail with index $\alpha_X$ rules out a Pareto consumption tail with
$\alpha_C>\alpha_X$.  The proof requires neither independence, homotheticity,
nor asymptotic linearity, and it does not assume that consumption itself has a
Pareto law.  The pointwise policy bound can also be weakened to a nonvanishing
exposure among households in the resource tail.  The observable converse is
equally useful.  If $\alpha_C>\alpha_X$, then among households with $X>x$, the
fraction consuming more than any fixed positive share of $X$ must converge to
zero (Corollary~\ref{cor:vanishing_exposure}).  The marginal tail comparison
therefore has a conditional policy diagnostic.

Two complementary sharpenings clarify the scope.  Feasibility supplies the
reverse bound: under the maintained positive lower bound, if consumption
also cannot exceed wealth and wealth has a Pareto-type tail, consumption
inherits the index as a weak Pareto-type tail, with arbitrary dependence
between wealth and the exogenous state (Corollary~\ref{cor:feas_inherit}).
Independence plus an exact Pareto wealth tail and an affine or concave
asymptotically linear policy deliver a Pareto tail for consumption with an
explicit constant (Proposition~\ref{th:2}).  The two results should not be
conflated: the first handles Markov dependence, while the second proves the
stronger distributional conclusion.

Third, Section~\ref{sec:ha_models} shows why the restriction binds in a
standard income-fluctuation model.  Condition~(19) of
\citet{ma2020income} implies $C\geq(1-\bar x)W$, while their cash-on-hand
choice set gives $C\leq W$.  Together, these inequalities imply that consumption inherits the weak
wealth-tail index whenever wealth has a Pareto-type tail, with arbitrary
Markov dependence (Corollary~\ref{cor:feas_inherit}).  The same stability certificate is maintained by the
model's stationarity and heavy-tail argument.  The route used to support a
stationary heavy wealth tail therefore prevents a thinner Pareto consumption
tail at the same time.  This circular dependence is the \emph{Pareto trap}.

A second class of applications is covered by the affine sufficient
condition.  In the finitely lived OLG economy of
\citet{benhabib2011distribution}, optimal consumption is affine in wealth
with state-dependent coefficients and a bounded human-wealth intercept, so
the lower bound holds even though consumption can exceed financial wealth,
and the model's own population power-tail bounds transfer to consumption
with the same exponent (Proposition~\ref{prop:olg}).
Appendix~\ref{app:survey} records, model by model, where the positive
linear lower bound is a theorem and where it remains conjectural.

The quantitative magnitude can be large.  For a full Pareto distribution with index $\alpha>1$, the
share held by the top $p$ fraction is $p^{1-1/\alpha}$.  Replacing
$\alpha_C=3.0$ by $\alpha_W=1.5$ raises the implied top 1\% consumption share
from $4.6\%$ to $21.5\%$, a factor of about $4.6$.  An empirical population share also depends on the contribution below the fitted threshold.  Section~\ref{sec:calibration}
uses calibrated policies and an explicitly auxiliary tail calculation to
illustrate the mechanism without treating a globally linearized wealth process
as a finite-wealth model fit.

The closest parametric comparison is \citet{gaillard2023consumption}.  In a
continuous-time economy, they derive complete mappings among consumption,
labor-income, wealth, and capital-income tails from asymptotic policy formulas
and an endogenous tail-factorization result.  In deriving the wealth tail in Proposition~3, they verify that the stationary joint density factors at high wealth into a Pareto term and a productivity term (their footnote~27), and Theorem~\ref{thm:3} here allows arbitrary dependence between wealth and the exogenous state.  The present result instead gives
a one-sided policy restriction and a discrete-time stability certificate.  It
does not claim to remove an independence assumption from, or subsume, their
framework.  The lower-bound argument also relates to the impossibility result
of \citet{stachurski2019impossibility}, under the corrected assumptions in
\citet{stachurski2020corrigendum}.  They show that canonical risk-free
Bewley--Huggett--Aiyagari models cannot generate wealth tails heavier than
income.  Here the same type of consumption lower bound is used in the other
direction: conditional on a heavy resource tail, it restricts consumption.
The paper thereby connects work on tail generation in heterogeneous-agent economies
\citep{benhabib2011distribution,benhabib2015wealth,
benhabib2018skewed,nirei2016pareto,jones2018schumpeterian,
gabaix2016dynamics,beare2022determination,achdou2022income} to evidence on
consumption and wealth inequality
\citep{krueger2006does,aguiar2015has,attanasio2016consumption,
meyer2017consumption}.

The scope is deliberately impossibility-focused.  The paper does not add a
structural escape mechanism and does not classify multi-asset or HANK models
without a model-specific lower-exposure certificate.  Prominent two-asset and
HANK benchmarks therefore remain outside the proved class rather than being
classified by analogy \citep{kaplan2014model,kaplan2018monetary}.  The proof-status tables in
Appendix~\ref{app:survey} separate formally established applications from
conjectural and open cases.  Section~\ref{sec:empirics}
presents the empirical tails, Section~\ref{sec:imp_key} proves the restriction
and its sharpenings, and Section~\ref{sec:ha_models} verifies the applications.
Section~\ref{sec:calibration} gives the quantitative diagnostic, and
Section~\ref{sec:recap} concludes.  Data and estimation details appear in
Appendix~\ref{app:empirics}.

\section{Empirical tails of wealth and consumption}
\label{sec:empirics}

Before turning to the estimates, we formalize the tail notation used in the introduction.

\begin{definition}[Pareto tail and tail index]
    \label{df:01}
    For a real-valued random variable $X$ with distribution $F_X$, write $\bar F_X(x) := \PP\{X > x\} = 1 - F_X(x)$ for the survival function, or complementary cumulative distribution function (CDF). The distribution $F_X$ has a \emph{Pareto tail} with index $\alpha>0$ if there exists $k>0$ such that
    \begin{equation}\label{eq:pareto_tail}
      \lim_{x \to \infty} x^{\alpha} \bar F_X(x) = k.
    \end{equation}
    A smaller index means a heavier tail. When the indices exist, $\alpha_C$ and $\alpha_W$ denote the consumption and wealth tail indices.
\end{definition}

This section estimates both tail indices of the wealth and consumption distributions in three household survey systems and fixes in advance which estimates may discipline the models of Sections~\ref{sec:ha_models} and~\ref{sec:calibration}. The data are the Household, Income and Labour Dynamics in Australia survey (HILDA, wave 22), the euro-area Household Finance and Consumption Survey (HFCS, wave 5, six countries), and the China Household Income Project (CHIP, 2013 and 2018). Each survey records consumption and wealth for the same households, so the two margins are estimated jointly rather than matched across sources. For the United States, \citet{gaillard2023consumption} document the same consumption and wealth ranking directly, and \citet{toda2015double} establish the power-law behavior of the consumption tail itself.

Consumption is each survey's annual household expenditure aggregate: an eighteen-component basket of nondurable goods and services in HILDA, expenditure on consumer goods and services in the HFCS, and total consumption expenditure in CHIP. The three concepts differ in breadth across the surveys rather than forming one nested definition, and the ordering reported below appears under each of them. 

Wealth is household net worth in HILDA and the HFCS. The CHIP resource is nonhousing assets net of matched liabilities, a narrower concept treated as a separate replication rather than a comparable net worth. HFCS values are converted to 2024 prices using each household's fieldwork year, while HILDA waves and CHIP years are estimated separately in source-year currency, so no cross-year price adjustment arises. 

{\setlength{\emergencystretch}{1em}Each index is a Pareto maximum likelihood estimate on strict exceedances above the fixed 90th percentile of that variable's positive observations, computed with provider survey weights in HILDA and the HFCS and with equal household weights in CHIP, and corrected for first-order finite-sample bias.\par}

Released HILDA wealth is top coded at official wave-specific cutoffs that overlap the fitted region, so the Australian wealth fit treats top-coded households as right-censored events \citep{wang2022tail}, with a truncated-window fit \citep{aban2006parameter} reported as the local description. In wave 22 the official cutoff is \$8{,}000{,}000 while the fitted tail begins at \$3{,}063{,}000, so the cutoff sits at 2.6 times the threshold, and the tail fit carries 729 exactly observed households plus 112 censored at the cutoff. 

We also hold access to the HILDA Restricted Release, in which wealth is not top coded, and use it below to validate the correction directly. Uncertainty is survey specific: HILDA uses 45 replicate weights, the HFCS supplies five implicates, that is, five completed versions of the survey data, each carrying 1{,}000 replicate weights, with estimates and covariance matrices combined across implicates by Rubin's rules \citep{rubin1987multiple}, and CHIP uses a paired bootstrap that resamples households separately within the urban and rural sectors. Appendix~\ref{app:empirics} records construction and estimator details.

Fitted indices support three claims of increasing strength: a description of the fitted range, an extrapolation to the tail beyond it, and a target a model may be calibrated to. The second and third require more than a good fit, so the design separates them in advance. The seven weighted-survey comparisons, Australia and the six HFCS countries, form one prespecified family for multiple testing. 

An index pair enters asymptotic comparisons only if it passes a four-part eligibility rule covering construction, fit, threshold stability, and inference, and ineligible pairs are retained in the family with a p-value of one, so screening cannot manufacture significance. The one-sided hypothesis that the consumption index exceeds the wealth index is tested with a Holm adjustment at the 5 percent familywise level \citep{holm1979simple}. The adjustment controls the probability of even one false rejection across the seven comparisons at the 5 percent level. Appendix~\ref{app:eligibility} defines the four checks and the step-down rule. The two Chinese samples form a separate family, reported over the observed range and not tested. A stricter exact-tail benchmark, which licenses calculations that assume a Pareto distribution outright, additionally requires eligible Pareto fits on both margins with both indices above one.

\begin{table}[H]
\centering
\footnotesize
\caption{Tail estimates and reporting status}
\label{ch2:tab:r2_tail_results}
\label{ch2:tab:tail_pooled_summary}
\resizebox{\textwidth}{!}{%
\begin{tabular}{llccc}
\toprule
Country and sample & Wealth/resource fit & $\widehat\alpha_C$ (s.e.) & $\widehat\alpha_W$ (s.e.) & $\widehat\alpha_C-\widehat\alpha_W$ (s.e.) \\
\midrule
Australia (HILDA), wave 22 & censored continuation & 3.168 (0.244) & 2.119 (0.159) & 1.050 (0.199)$^{*}$ \\
 & local truncated (u,c) & 3.168 (0.244)$^{\ddagger}$ & 2.027 (0.308)$^{\ddagger}$ & 1.141 (0.242) \\
\midrule
Germany (HFCS), wave 5 & uncensored working fit & 4.085 (0.220) & 1.603 (0.132) & 2.482 (0.276)$^{*}$ \\
Spain (HFCS), wave 5 & uncensored working fit & 3.076 (0.166)$^{\dagger}$ & 1.659 (0.080)$^{\dagger}$ & 1.417 (0.188) \\
Finland (HFCS), wave 5 & uncensored working fit & 4.685 (0.126)$^{\dagger}$ & 1.856 (0.057)$^{\dagger}$ & 2.830 (0.139) \\
France (HFCS), wave 5 & uncensored working fit & 2.740 (0.109)$^{\dagger}$ & 1.726 (0.060)$^{\dagger}$ & 1.014 (0.121) \\
Italy (HFCS), wave 5 & uncensored working fit & 3.383 (0.214) & 1.615 (0.127) & 1.768 (0.223)$^{*}$ \\
Netherlands (HFCS), wave 5 & uncensored working fit & 2.906 (0.235) & 2.096 (0.132) & 0.810 (0.271) \\
\midrule
China (CHIP), 2013 & nonhousing resource & 2.830 (0.093)$^{\ddagger}$ & 1.818 (0.069)$^{\ddagger}$ & 1.012 (0.108) \\
China (CHIP), 2018 & nonhousing resource & 2.808 (0.057)$^{\ddagger}$ & 1.613 (0.036)$^{\ddagger}$ & 1.195 (0.065) \\
\bottomrule
\end{tabular}
}
\par\vspace{4pt}
{\footnotesize $^{*}$ marks rejection of a non-positive index difference
after the fixed-family Holm adjustment defined in the text. $^{\dagger}$
marks an index pair not qualified under the four-part eligibility rule. $^{\ddagger}$
marks a window-local fit with no asymptotic index claim. Markers qualify the
index pair or row, not either marginal fit in isolation. The unmarked Dutch
pair is eligible, but its adjusted one-sided test does not reject a
non-positive difference. The Chinese samples form a separate family and are
not tested.}
\end{table}

Table~\ref{ch2:tab:r2_tail_results} reports the estimates and their reporting status. Four index pairs are eligible: Australia, Germany, Italy, and the Netherlands. The fitted consumption index exceeds the fitted wealth index in all seven weighted-survey comparisons, and the difference survives the Holm adjustment for Australia (gap \HildaWaveTwentyTwoGap{}, standard error \HildaWaveTwentyTwoSEGap{}), Germany (\HfcsDeGap{}, \HfcsDeSEGap{}), and Italy (\HfcsItGap{}, \HfcsItSEGap{}). The Dutch pair is eligible, and its adjusted test does not reject (adjusted p-value 0.081). Spain, Finland, and France show positive fitted differences but fail eligibility, so their rows are descriptive. The Chinese comparisons are positive in both years (\ChipsTwentyThirteenGap{}, standard error \ChipsTwentyThirteenSEGap{}, in 2013 and \ChipsTwentyEighteenGap{}, \ChipsTwentyEighteenSEGap{}, in 2018) over the observed range only. Only Germany and Italy pass the exact-tail benchmark.

\begin{figure}[tbp]
    \centering
    \begin{tikzpicture}
\fill[gray!25] (0.000,3.868) -- (0.213,4.079) -- (0.433,4.307) -- (0.662,3.715) -- (0.899,3.521) -- (1.145,3.421) -- (1.400,3.621) -- (1.664,3.353) -- (1.938,4.166) -- (2.223,3.178) -- (2.518,3.400) -- (2.824,3.630) -- (3.141,3.057) -- (3.470,2.990) -- (3.811,2.958) -- (4.165,2.812) -- (4.531,3.069) -- (4.912,2.735) -- (5.306,2.731) -- (5.716,2.505) -- (6.140,2.478) -- (6.580,3.071) -- (7.036,2.461) -- (7.509,2.464) -- (8.000,2.740) -- (8.000,0.821) -- (7.509,0.920) -- (7.036,0.898) -- (6.580,0.845) -- (6.140,0.999) -- (5.716,1.059) -- (5.306,1.134) -- (4.912,1.203) -- (4.531,1.162) -- (4.165,1.407) -- (3.811,1.474) -- (3.470,1.623) -- (3.141,1.905) -- (2.824,1.611) -- (2.518,1.877) -- (2.223,2.180) -- (1.938,1.630) -- (1.664,2.040) -- (1.400,2.337) -- (1.145,2.472) -- (0.899,2.653) -- (0.662,2.846) -- (0.433,2.789) -- (0.213,3.309) -- (0.000,3.868) -- cycle;
\draw[->] (0,0) -- (8.500,0) node[right] {$t$};
\draw[->] (0,0) -- (0,4.900) node[above] {$R_{0.90}(t)$};
\draw (0.000,0.06) -- (0.000,-0.06) node[below] {1};
\draw (2.870,0.06) -- (2.870,-0.06) node[below] {1.5};
\draw (5.740,0.06) -- (5.740,-0.06) node[below] {2};
\draw (0.06,1.934) -- (-0.06,1.934) node[left] {0.5};
\draw (0.06,3.868) -- (-0.06,3.868) node[left] {1};
\draw[dotted] (0,3.868) -- (8.000,3.868);
\draw[very thick] (0.000,3.868) -- (0.213,3.674) -- (0.433,3.466) -- (0.662,3.252) -- (0.899,3.056) -- (1.145,2.908) -- (1.400,2.909) -- (1.664,2.615) -- (1.938,2.606) -- (2.223,2.632) -- (2.518,2.526) -- (2.824,2.418) -- (3.141,2.413) -- (3.470,2.203) -- (3.811,2.088) -- (4.165,1.989) -- (4.531,1.888) -- (4.912,1.814) -- (5.306,1.760) -- (5.716,1.629) -- (6.140,1.573) -- (6.580,1.610) -- (7.036,1.487) -- (7.509,1.505) -- (8.000,1.500);
\node[right] at (8.000,1.500) {ratio};
\end{tikzpicture}
    \caption{Survival-order profile for the Australian baseline (HILDA wave 22). The curve is the ratio of the two normalized survival functions above their own weighted 90th percentile anchors $v_C$ and $v_W$,
    $R_{0.90}(t) = \dfrac{\PP\{C \ge t\,v_C\}/\PP\{C \ge v_C\}}{\PP\{W \ge t\,v_W\}/\PP\{W \ge v_W\}}$,
    plotted against the multiplier $t$. The shaded region is the released simultaneous 95 percent band for this profile. The anchors are taken on the full matched sample, so the wealth anchor, \$3{,}020{,}000, sits slightly below the \$3{,}063{,}000 positive-sample threshold of the index fit.}
    \label{fig:survival_order}
\end{figure}
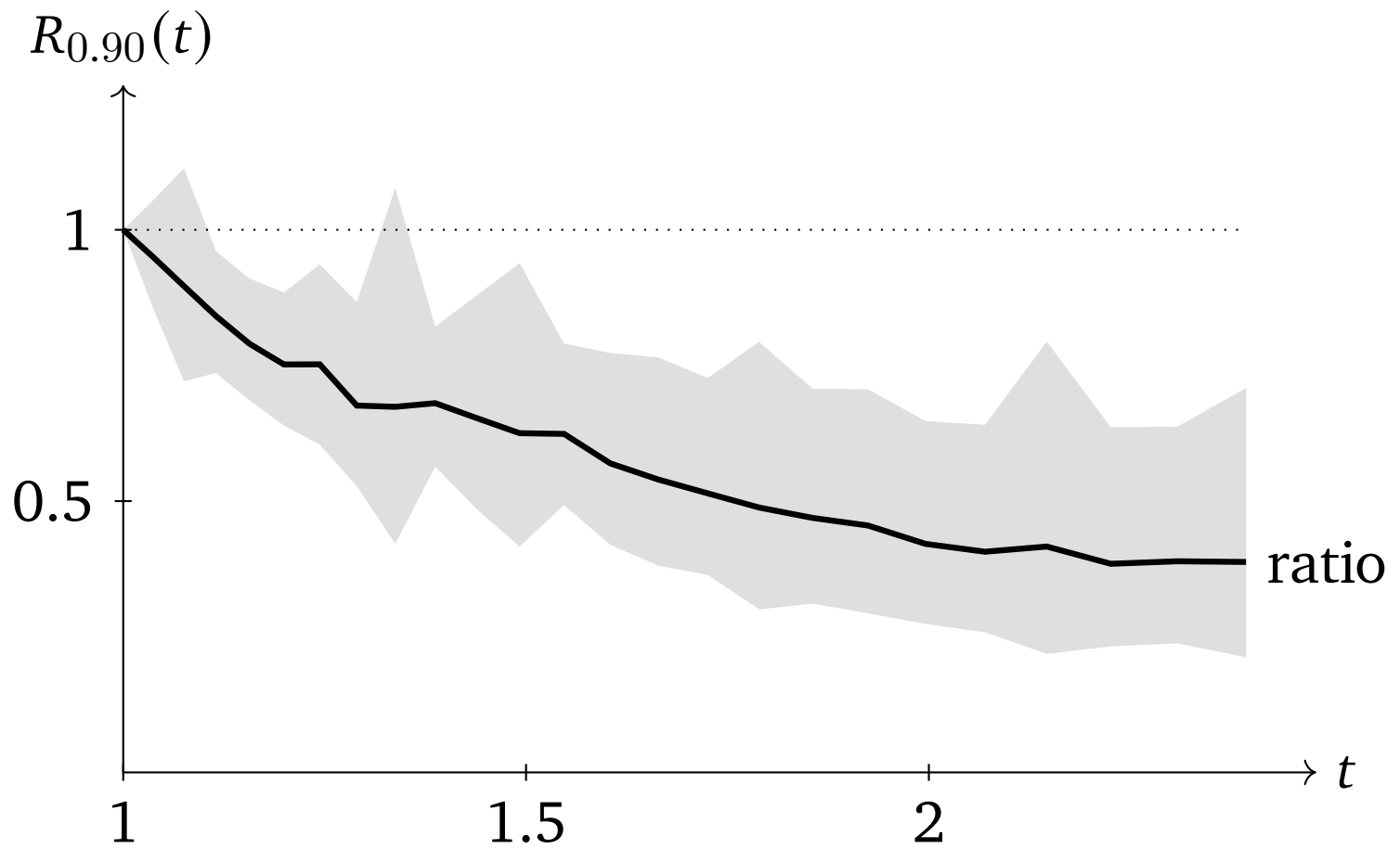

Figure~\ref{fig:survival_order} shows the Australian profile. The estimated ratio is below one at every plotted multiplier above the anchor. At the largest supported multiplier, 2.394, it is 0.388, with a simultaneous 95 percent band of 0.212 to 0.708.

Two further checks support the Australian correction and the German estimate. Rerunning the identical estimation procedure on the HILDA Restricted Release, where wealth is not top coded, reproduces the corrected wealth indices to within about 0.05 to 0.07 index points in each of the five matched waves, with the restricted estimate the slightly larger of the two every time, and no wave rejects equality of the pair after the prespecified two-sided Holm adjustment across the five waves. The matched results support the correction, and failure to reject is not proof of equivalence. Independently, \citet{koenig2025routes} obtain a German wealth index of 1.66 from the 2019 German Socio-Economic Panel including its high-wealth household sample, close to the \HfcsDeAlphaResource{} estimated here.

The four eligible pairs are the admissible model targets of this paper. The calibration discipline is fixed here: each eligible wealth index supplies a separate calibration target, and the model's consumption index is an untargeted prediction for comparison with the corresponding survey estimate. The evidence is a contemporary baseline: coverage spans eight countries, four samples pass the prespecified checks, and the earlier HFCS wave shows a different eligibility pattern, so the claim is about the current waves, not about every country in every wave.

\section{The impossibility theorem}
\label{sec:imp_key}

This section states the paper's central restriction. If stationary wealth has a Pareto-type tail and consumption is bounded below by a positive fraction of wealth at high wealth (Assumption~\ref{a:1}), then stationary consumption cannot have a Pareto-type tail index strictly larger than that of wealth (Theorem~\ref{thm:3}). The economic work is therefore to identify conditions under which optimal policies satisfy this bound.

The simplest case illustrates the logic: if optimal consumption is a fixed fraction of wealth, independent of the exogenous state, the two distributions share the same tail index (Lemma~\ref{lem:power_transform} with $\rho=1$). More generally, an exact affine policy or a concave asymptotically linear policy, each conditional on the exogenous state, implies Assumption~\ref{a:1} (Proposition~\ref{prop:00}). If wealth has a Pareto tail, not merely a Pareto-type tail, and is independent of the exogenous state, then consumption has a Pareto tail with the same index and an explicit constant (Proposition~\ref{th:2}).

\subsection{Tail classes}
\label{ssec:tail_classes}

Definition~\ref{df:01} fixed the Pareto tail through the limit $x^{\alpha}\bar F_X(x)\to k$. The theorem needs only the weaker requirement that the survival function behaves like a power under rescaling, and its corollaries use a weaker logarithmic index still. This subsection fixes both classes \citep{bingham1987regular}.

\begin{definition}[Regular variation]
\label{def:reg_var}
A measurable function $J:(0,\infty)\to(0,\infty)$ is \emph{regularly varying} with index $\eta\in\RR$, written $J\in RV_\eta$, if
\[
        \frac{J(tx)}{J(x)} \to t^\eta
        \qquad \text{as } x\to\infty
        \quad \text{for every fixed } t>0 .
\]
When $\eta=0$, $J$ is said to be \emph{slowly varying}.
\end{definition}

\begin{definition}[Pareto-type and weak Pareto-type tails]
\label{def:pttail}
Let $X\geq 0$ and let $\alpha>0$.
\begin{enumerate}[label=(\roman*)]
    \item $X$ has a \emph{Pareto-type tail} with index $\alpha$ if
    there exist $x_0>0$ and a slowly varying function $L$ such that
    $\bar F_X(x)=x^{-\alpha}L(x)$ for every $x\geq x_0$.
    Equivalently, $\bar F_X\in RV_{-\alpha}$.
    \item $X$ has a \emph{weak Pareto-type tail} with index $\alpha$
    if there exists $x_0>0$ such that $\bar F_X(x)>0$ for all
    $x\geq x_0$, and $-\log\bar F_X(x)/\log x\to\alpha$.
    \item $X$ has a \emph{Pareto distribution} with index $\alpha$
    and threshold $x_{\min}>0$ if
    $\bar F_X(x)=(x_{\min}/x)^\alpha$ for all $x\geq x_{\min}$.
\end{enumerate}
In each case, $\alpha$ is the tail index.
\end{definition}

\begin{sloppypar}
The hierarchy is: Pareto distribution implies Pareto tail implies Pareto-type tail implies weak Pareto-type tail. Under a Pareto distribution, $x^\alpha\bar F_X(x)=x_{\min}^\alpha$ for every $x\geq x_{\min}$, so the tail constant is $k_X=x_{\min}^\alpha$, and the distinction matters for top-share calculations: the top-share identity used in Section~\ref{sec:calibration} is exact only for the full Pareto distribution. A Pareto tail is Pareto-type, because every positive function converging to a positive constant is slowly varying, and the converse fails: $\bar F_X(x)=x^{-\alpha}\log x$ for large $x$ is Pareto-type but not a Pareto tail. Every Pareto-type tail is weak Pareto-type with the same index, because $\log L(x)/\log x\to0$ for every slowly varying $L$ \citep[Proposition~1.3.6]{bingham1987regular}. The converse again fails: $\bar F_X(x)=(2+\sin(\log x))/(3x)$ for $x\geq1$ is squeezed between $1/(3x)$ and $1/x$, so its logarithmic index is one, yet the ratio $\bar F_X(e^{\pi}x)/\bar F_X(x)$ oscillates, so $\bar F_X\notin RV_{-1}$. The weak class takes the Pareto-exponent criterion of \citet[Theorem~1]{beare2022determination} as the definition. It coincides with the weak Pareto type of \citet[Chapter~7]{gulisashvili2012regularly} and is weaker than the weak Pareto law of \citet[Section~1.2]{mandelbrot1960pareto}, which requires $\bar F_X(x)\sim kx^{-\alpha}$.
\end{sloppypar}

If $X$ and $Y$ have Pareto-type tails with $\alpha_X>\alpha_Y$, then the ratio $\bar F_X(x)/\bar F_Y(x)$ equals $x^{-(\alpha_X-\alpha_Y)}L_X(x)/L_Y(x)$, which tends to zero because every positive power dominates a slowly varying function \citep[Proposition~1.3.6]{bingham1987regular}. Within these classes we therefore describe the variable with the larger index as having the thinner tail.

\subsection{Statement}
\label{sec:stim}

Let $(\Omega, \mathcal F, \mathbb P)$ be the probability space on which all random variables are defined. Let $W\in\RR_+$ denote stationary wealth, let $S$ be an exogenous random element of a measurable state space $\mathbb S$, and let $C=f(W,S)$ denote stationary consumption, where $f:\RR_+\times\mathbb S\to\RR_+$ is a measurable consumption policy. The results of this section use only the stated conditions on $f$ and not the household problem behind them. Optimality enters in Section~\ref{sec:ha_models}, which derives the bound from the income-fluctuation problem.

\begin{assumption}[Positive linear lower bound]
    \label{a:1}
    There exist constants $a>0$ and $w_0 \geq 0$ such that
    \begin{equation}
    \label{eq:a1}
        f(w, s) \geq a  w
    \end{equation}
    for all $w \geq w_0$ and all $s \in \mathbb S$.
\end{assumption}

\begin{remark}
	Assumption~\ref{a:1} is the key condition in the theorem. It says that beyond some wealth level, consumption never falls below a fixed fraction $a$ of wealth. The lower bound need not hold globally (only eventually, for $w \geq w_0$) and the policy need not be linear.
\end{remark}

\begin{theorem}[Impossibility of a thinner consumption tail under a positive linear lower bound]
\label{thm:3}
Let $W,S,C,f$ be defined as above and suppose $W$ has a
Pareto-type tail with index $\alpha_W>0$. If Assumption~\ref{a:1} holds, then
\[
    \liminf_{x\to\infty}
    \frac{\bar F_C(x)}{\bar F_W(x)}
    \geq a^{\alpha_W}>0.
\]
If, in addition, $C$ has a Pareto-type tail with index $\alpha_C$,
then $\alpha_C\leq\alpha_W$.
\end{theorem}

\begin{proof}
Let $a$ and $w_0$ satisfy Assumption~\ref{a:1}. If $x\geq aw_0$ and
$W>x/a$, then $W\geq w_0$ and $C=f(W,S)\geq aW>x$. Hence, for all
$x\geq aw_0$,
$\bar F_C(x)=\PP\{C>x\}\geq\PP\{W>x/a\}=\bar F_W(x/a)$.
Since $\bar F_W\in RV_{-\alpha_W}$,
$\bar F_W(x/a)/\bar F_W(x)\to a^{\alpha_W}$.
This gives the liminf bound, so $\bar F_C(x)/\bar F_W(x)$ does not
converge to zero. Finally, suppose $C$ has a Pareto-type tail with
index $\alpha_C$, hence a weak Pareto-type tail with the same index
(Subsection~\ref{ssec:tail_classes}). The same lower bound gives
$-\log\bar F_C(x)/\log x\leq-\log\bar F_W(x/a)/\log x$
for all sufficiently large $x$. The left side tends to
$\alpha_C$, while the right side tends to $\alpha_W$. Hence
$\alpha_C\leq\alpha_W$.
\end{proof}

\begin{remark}[A weaker tail-exposure condition]
\label{rem:tail_exposure}
The uniform policy bound is a transparent sufficient condition, not the weakest possible one. If, for some $a>0$,
\begin{equation}
\label{eq:tail_exposure}
    p:=\liminf_{x\to\infty}
    \PP\{C\geq aW\mid W>x\}>0,
\end{equation}
then
\[
    \liminf_{x\to\infty}\frac{\bar F_C(x)}{\bar F_W(x)}\geq p\,a^{\alpha_W}>0,
\]
and $\alpha_C\leq\alpha_W$ whenever $C$ has a Pareto-type tail. The proof is unchanged apart from the constant: $\bar F_C(x)\geq p'\,\bar F_W(x/a)$ for every $p'\in(0,p)$ and all sufficiently large $x$, and a positive constant does not affect the logarithmic limits. An independent Bernoulli exposure with success probability $p$ attains the smaller coefficient, so the constant is sharp in general. Assumption~\ref{a:1} implies \eqref{eq:tail_exposure} with conditional probability one.
\end{remark}

\subsection{Intuition: the linear case}
\label{sec:explain_impossibility}

A simple counterfactual makes the logic concrete. In the
Australian data, the complementary CDF of nondurable consumption falls
more steeply on logarithmic axes than that of net worth
($\widehat\alpha_C=\HildaWaveTwentyTwoAlphaConsumption$ against
$\widehat\alpha_W=\HildaWaveTwentyTwoAlphaResource$,
Table~\ref{ch2:tab:r2_tail_results}). A counterfactual rule
$C=0.5W$ shifts the wealth tail horizontally but does not change its
slope. This is exactly what Lemma~\ref{lem:power_transform}
predicts at $\rho=1$: changing the constant $a$ shifts the counterfactual line
but does not alter its tail index. A linear consumption rule forces
$\alpha_C = \alpha_W$, ruling out the empirical ranking
$\alpha_C > \alpha_W$. Extending the argument to state-dependent
policies requires the conditions of Subsection~\ref{ssec:suff}.

Consider households in a Pareto-type wealth tail, regardless of how that
tail is generated. If the policy keeps consumption above a fixed positive
fraction of wealth at high wealth levels, then exceptionally wealthy
households must also have exceptionally high consumption. The consumption
tail is therefore at least as heavy as the wealth tail.

\subsection{Sufficient conditions: affine and asymptotically linear policies}
\label{ssec:suff}

The sufficient conditions in this subsection use compactness of the
state space and continuity of the policy to obtain a uniform positive
coefficient.

Recall from Definition~\ref{df:asym_linear}(i) that a function
$\varphi:\RR_+\to\RR_+$ is asymptotically linear if $\varphi(x)/x\to k$ for
some constant $k>0$. Here the definition is applied state by state, to
$w\mapsto f(w,s)$ for each $s\in\mathbb S$.

\begin{assumption}[State space]
    \label{a:space}
    The state space $\mathbb S$ is a compact metric space equipped with
    its Borel sigma-algebra.
\end{assumption}

\begin{assumption}[Exact affine policy]
    \label{a:3}
    Suppose Assumption~\ref{a:space} holds. There exist continuous
    functions
    $h: \mathbb S \to (0, \infty)$ and
    $g: \mathbb S \to [0, \infty)$ such that
    $f(w,s)=h(s)w+g(s)$ for all
    $(w,s)\in\RR_+\times\mathbb S$.
\end{assumption}

\begin{assumption}[Concave asymptotically linear policy]
    \label{a:5}
    Suppose Assumption~\ref{a:space} holds and $f$ is continuous on
    $\RR_+\times\mathbb S$. For every $s\in\mathbb S$, the function
    $f(\cdot,s)$ is concave. There exists a continuous
    function $K: \mathbb S \to (0, \infty)$ such that
    \begin{equation}
        \label{eq:a3}
        \lim_{w \to \infty} f(w, s)/w = K(s)
    \end{equation}
    for all $s \in \mathbb S$.
\end{assumption}

\begin{proposition}
	\label{prop:00}
	Let $f$ be defined as above. If $f$ satisfies either Assumption~\ref{a:3} or Assumption~\ref{a:5},
	then $f$ satisfies Assumption~\ref{a:1}.
\end{proposition}

\begin{proof}[Proof of Proposition~\ref{prop:00}]
	Case 1: Suppose $f$ satisfies Assumption~\ref{a:3}. Since $h$ is continuous on the compact set $\mathbb S$ with $h:\mathbb S \to (0, \infty)$, the minimum $a:= \min_{s \in \mathbb S} h(s)$ is positive. Then $f(w, s) = h(s) w + g(s) \geq aw$ for all $w \in \RR_+$ and all $s \in \mathbb S$. Thus Assumption~\ref{a:1} holds with $w_0 = 0$.

	Case 2: Suppose $f$ satisfies Assumption~\ref{a:5}. Fix $s \in \mathbb S$. Since $f(\cdot,s)$ is concave with $f(0,s) \geq 0$, the ratio $f(w,s)/w$ is nonincreasing in $w$ on $(0,\infty)$: for $0 < w_1 < w_2$, write $w_1 = (w_1/w_2)\,w_2 + (1 - w_1/w_2)\cdot 0$, so concavity gives $f(w_1,s) \geq (w_1/w_2)\,f(w_2,s)$, hence $f(w_1,s)/w_1 \geq f(w_2,s)/w_2$. It follows that $f(w,s)/w \;\geq\; \lim_{w'\to\infty} f(w',s)/w' \;=\; K(s)$ for all $w > 0$. Since $K$ is continuous on the compact set $\mathbb S$ with $K(s) > 0$, setting $a := \min_{s\in\mathbb S} K(s) > 0$ gives $f(w,s) \geq a\,w$ for all $w > 0$ and all $s \in \mathbb S$. Thus Assumption~\ref{a:1} holds with $w_0 = 0$.
\end{proof}

\begin{remark}
		Both sufficient conditions deliver the global bound with $w_0 = 0$. This is stronger than the tail condition required by Theorem~\ref{thm:3}.
\end{remark}

\subsection{Exact tail index inheritance under independence}

When wealth has a Pareto tail, not merely a Pareto-type tail, and is independent of the exogenous state, the affine and concave asymptotically linear policies of Subsection~\ref{ssec:suff} transmit a Pareto tail with the same index and an explicit constant.

\begin{proposition}[Exact tail inheritance]
	\label{th:2}
    Let $W,S,C$ be defined as above, and suppose $W$ has a Pareto tail
    with index $\alpha_W>0$ and constant $k_W$. If either
    Assumption~\ref{a:3} or~\ref{a:5} holds and $W$ and $S$
    are independent, then $C$ has a Pareto tail with the same index.
    Its constant is
    \[
        k_C=k_W\EE[H(S)^{\alpha_W}],
    \]
    where $H=h$ under Assumption~\ref{a:3} and $H=K$ under
    Assumption~\ref{a:5}.
\end{proposition}

\begin{proof}[Sketch of the proof]
\sloppy
Under Assumption~\ref{a:3}, boundedness of $g$ on compact
$\mathbb S$ gives uniform convergence of $f(w,s)/w$ to $h(s)$. Under
Assumption~\ref{a:5}, concavity makes $f(w,s)/w$ decrease to
$K(s)$. Dini's theorem \citep[Theorem~7.13]{rudin1976principles} then makes
the convergence uniform on $\mathbb S$. In either case, $f$ has uniform asymptotic power growth
with exponent one. Corollary~\ref{cor:exact_tail_mapping} gives the
result. Appendix~\ref{app:imp_eta_proof} supplies the uniformity
details.
\end{proof}

\begin{remark}
\sloppy
\citet[Proposition~1]{gaillard2023consumption} obtain
$\alpha_C=\alpha_W$ in their continuous-time model with homothetic
preferences and scale-independent returns, under their
Proposition~1 conditions.
Proposition~\ref{th:2} establishes the same equality under its
policy conditions and independence, whereas Theorem~\ref{thm:3}
rules out $\alpha_C>\alpha_W$ under its stated assumptions, without
independence or a tail-factorization condition.
Appendix~\ref{app:discussion} compares these assumptions and states
the open extension.
\end{remark}

\subsection{The escape criterion}
\label{ssec:escape_criterion}

Theorem~\ref{thm:3} applies whenever wealth has a Pareto-type tail
and consumption satisfies Assumption~\ref{a:1}, so a model matching
the empirical ranking $\alpha_C>\alpha_W$ must break the positive linear
lower bound. Equivalently,
\[
  \liminf_{w\to\infty}\inf_{s\in\mathbb S}\frac{f(w,s)}{w}=0,
\]
which is weaker than requiring $f(w,s)/w\to0$ in every state. The
ranking also has a distributional counterpart.

\begin{corollary}[Vanishing exposure in the wealth tail]
\label{cor:vanishing_exposure}
If $W$ and $C$ have Pareto-type tails with $\alpha_C>\alpha_W$, then, for every
$a>0$,
\begin{equation}
\label{eq:vanishing_exposure}
    \PP\{C>aW\mid W>x\}\longrightarrow0
    \qquad\text{as }x\to\infty.
\end{equation}
\end{corollary}

\begin{proof}
Since $\{C>aW,\,W>x\}\subseteq\{C>ax\}$,
\[
 \PP\{C>aW\mid W>x\}
 \leq\frac{\bar F_C(ax)}{\bar F_W(x)}
 =\frac{\bar F_C(ax)}{\bar F_C(x)}
  \cdot\frac{\bar F_C(x)}{\bar F_W(x)}.
\]
The first factor converges to $a^{-\alpha_C}$ because
$\bar F_C\in RV_{-\alpha_C}$, and the second tends to zero because
$\alpha_C>\alpha_W$ (Subsection~\ref{ssec:tail_classes}).
\end{proof}

Among increasingly wealthy households, the fraction consuming any fixed positive share of wealth must therefore vanish. This is necessary for the observed tail ranking, not sufficient to generate it: $C=W/\log(e+W)$ has a vanishing share but the same Pareto-type index as $W$. The paper uses this statement as a diagnostic rather than developing a structural escape mechanism.

One sufficient condition for escape is uniform asymptotic power growth with policy exponent
$\rho\in(0,1)$ in the sense of Definition~\ref{def:policy_growth}.
Then $f(w,s)/w\to0$ uniformly in $s$. If wealth has a Pareto-type tail
and the remaining conditions of Theorem~\ref{thm:tail_mapping} hold,
then $C$ has a weak Pareto-type tail (Definition~\ref{def:pttail}(ii))
with index $\alpha_C=\alpha_W/\rho$.
Corollary~\ref{cor:exact_tail_mapping} gives a Pareto tail under its
stronger conditions.
Under a stylized wealth-only selection model
\citep[Chapter~2]{hu2026pareto}, both observed indices are inflated but
their ratio is unchanged. This invariance need not hold when selection
depends on consumption or state, the scale factor is unbounded, or frame or
measurement error is present.

\section{The income-fluctuation model}
\label{sec:ha_models}

The income-fluctuation model of \citet{ma2020income} extends
the canonical Bewley--Huggett--Aiyagari environment to
Markov-modulated discount factors, returns, and income. This section
states the problem in full and shows that a stability condition on
the savings rate (Assumption~\ref{ass:stability}) forces the
positive linear lower bound of Assumption~\ref{a:1} for any
admissible utility \citep[Proposition~2.6]{ma2020income}, so
Theorem~\ref{thm:3} applies.

Under CRRA utility, the same stability condition is among the
conditions that generate heavy-tailed wealth through a Kesten-type
argument \citep[Theorem~3.3]{ma2020income}
\citep{kesten1973random,goldie1991implicit}, creating the
Pareto trap (Theorem~\ref{thm:trap}). Shifted
HARA utility delivers the same bound under the alternative
foundation of \citet{ma2022asymptotic}
(Corollary~\ref{cor:trap_hara}).

\subsection{The income-fluctuation problem}
\label{sec:applic01}

We adopt the setup of \citet{ma2020income}. Let
$\mathbb S=\{s_1,\ldots,s_n\}$ be a finite exogenous state space and
let $\{s_t\}$ be an irreducible Markov chain with transition matrix
$Q=(Q_{ij})$. Irreducibility gives the finite exogenous chain a unique
stationary distribution and makes every state reachable from every other
state. Each household has wealth
$w_t\in\RR_+$ and state $s_t\in\mathbb S$, and chooses
$c_t\in[0,w_t]$. Set $\beta_0:=1$. The objective is
\begin{equation}\label{eq:objective}
  \mathbb E_0 \sum_{t=0}^{\infty} \left(\prod_{i=0}^{t} \beta_i\right) u(c_t)
\end{equation}
subject to
\begin{equation}\label{eq:budget}
  w_{t+1} = R_{t+1}(w_t - c_t) + Y_{t+1}, \qquad 0 \leq c_t \leq w_t,
\end{equation}
where $u:\RR_+\to\RR\cup\{-\infty\}$ is the utility function. Here
$w_t$ is wealth available before consumption, so $w_t-c_t\geq0$ is
saving carried into the return process. The feasibility restriction
therefore combines the timing convention of \citet{ma2020income}
with a nonnegative-saving constraint. Some Bewley--Huggett--Aiyagari
formulations instead use beginning-of-period assets and a separate
lower bound on next-period assets, which may permit negative
positions. The restriction $c_t\leq w_t$
enters the irreducibility and ergodicity arguments because the
borrowing constraint must bind with positive probability
\citep[Lemmas~C.2--C.3]{ma2020income}.

\begin{remark}
The constraint $c_t\leq w_t$ is a primitive of the environment, not a
derived property. Under the conditions of
Corollary~\ref{cor:trap_crra} it cannot bind asymptotically:
$m_\infty(s)<1$ in every state, so consumption falls strictly inside
the budget set at high wealth.
\end{remark}

For $t\geq1$, the discount factor, gross return, and nonfinancial income
are
\[
  \beta_t := \beta(s_t,\varepsilon_t), \qquad R_t := R(s_t,\zeta_t), \qquad Y_t := Y(s_t,\eta_t),
\]
for measurable functions $\beta:\mathbb S\times\RR\to(0,\infty)$,
$R:\mathbb S\times\RR\to(0,\infty)$, and
$Y:\mathbb S\times\RR\to\RR_+$. The innovation sequences are i.i.d.\
across time, mutually independent, and independent of $\{s_t\}$.
Let $\mathbb X:=\RR_+\times\mathbb S$. The initial
condition $(w_0,s_0)\in\mathbb X$ is given, and $\mathbb E_0$ denotes
expectation conditional on it.

\subsubsection{Existence and uniqueness of the optimal policy}

The following conditions ensure existence and uniqueness of an
optimal policy \citep[Sections~2.1--2.4]{ma2020income}:

\begin{assumption}[Existence and uniqueness conditions]
\label{ass:existence}
\begin{enumerate}[label=(\roman*)]
  \item \textbf{Utility.} $u$ is twice differentiable on $(0,\infty)$, with $u' > 0$ and $u'' < 0$. Moreover, $\lim_{c \to 0^{+}} u'(c) = \infty$ and $\lim_{c \to \infty} u'(c) < 1$.
  \item \textbf{Discounting.} $\lim_{n \to \infty} (\EE\prod_{t=1}^{n} \beta_t)^{1/n} < 1$.
  \item \textbf{Growth.} $\lim_{n \to \infty} (\EE\prod_{t=1}^{n} \beta_t R_t)^{1/n} < 1$.
  \item \textbf{Integrability.} $\EE Y < \infty$ and $\EE u'(Y) < \infty$.
\end{enumerate}
\end{assumption}

Condition~(i) imposes strict concavity, the Inada condition at zero,
and the high-consumption bound on marginal utility used by
\citet{ma2020income}. The leading
example is CRRA utility, $u(c) = c^{1-\gamma}/(1-\gamma)$ for
$\gamma > 0$ and $\gamma\neq1$, with $u(c)=\ln c$ when $\gamma=1$.
Condition~(ii) requires geometric discounting on average.
Condition~(iii) bounds the long-run discounted return, preventing
indefinite postponement of consumption. Condition~(iv) requires
integrability of income and of marginal utility evaluated at income.

Under Assumption~\ref{ass:existence}, the time-iteration operator of
\citet[Theorem~2.2]{ma2020income} has a unique fixed point $f$ in their
class of continuous, wealth-increasing feasible policies, and this fixed
point is the class's unique optimal policy. Extend it to zero wealth by
setting $f(0,s)=0$. Then $f:\mathbb X\to\RR_+$ satisfies
$0\leq f(w,s)\leq w$. The optimal saving policy
$g(w,s):=w-f(w,s)$ is also increasing in wealth
\citep[Proposition~2.3]{ma2020income}.

Write $\beta':=\beta(s',\varepsilon')$, $R':=R(s',\zeta')$, and
$Y':=Y(s',\eta')$ for next-period variables, and let $\EE_s$ denote
expectation conditional on current state $s$. The value function
satisfies the Bellman equation
\begin{equation}
\label{eq:bellman_operator}
    V(w,s)=
    \max_{0\leq c\leq w}
    \left\{
        u(c)+
        \EE_s\!\left[
        \beta'\, V\bigl(R'(w-c)+Y',s'\bigr)
        \right]
    \right\}.
\end{equation}
At interior states where the relevant derivatives exist, the Euler
equation takes the form
\begin{equation}
\label{eq:euler_policy}
    u'\bigl(f(w,s)\bigr)
    =
    \EE_s\!\left[
        \beta' R'\,
        u'\bigl(f(R'(w-f(w,s))+Y',s')\bigr)
    \right],
\end{equation}
and the envelope identity is
\begin{equation}
\label{eq:envelope_consumption}
    V_w(w,s)=u'\bigl(f(w,s)\bigr).
\end{equation}
These equations are used only to verify the high-wealth shape of
$f$.

\subsubsection{The stability condition}

The stability condition serves four roles: it forces optimal consumption to satisfy Assumption~\ref{a:1} (Proposition~\ref{prop:stab}), it provides the savings upper bound for stationarity of the wealth distribution \citep[Theorem~3.1]{ma2020income}, it supports uniqueness and ergodicity of that distribution \citep[Theorem~3.2]{ma2020income}, and it enters the conditions used to obtain a Pareto-order lower bound for wealth \citep[Theorem~3.3]{ma2020income}.

\begin{assumption}[Stability condition]\label{ass:stability}
There exists $\bar{x} \in (0,1)$ such that, for every $w>0$ and
$i=1,\ldots,n$,
\begin{equation}\label{eq:euler_ineq}
  \sum_{j=1}^{n} Q_{ij}\; \EE\!\left[\beta(s_j,\varepsilon)\, R(s_j,\zeta)\, u'\!\bigl(R(s_j,\zeta)\,\bar{x}\,w\bigr)\right] \;\leq\; u'(w).
\end{equation}
\end{assumption}

Conditional on current state $s_i$, the left-hand side of
\eqref{eq:euler_ineq} is the expected marginal value of saving
fraction $\bar{x}$ of wealth $w$ when nonfinancial income is zero. The
right-hand side is the marginal utility of consuming all wealth. The
inequality says saving fraction $\bar{x}$ is suboptimal even in this
zero-income case, forcing $f(w,s) \geq (1-\bar{x})w > 0$ for all
$w > 0$ and $s \in \mathbb S$ (Proposition~\ref{prop:stab}). This
is \citet[condition~(19)]{ma2020income}.

\begin{remark}[Income and the stability condition]
\label{rem:stability_income}
Since $Y \geq 0$ and $u'' < 0$, adding income $Y(s_j,\eta)$ inside $u'(\cdot)$ in~\eqref{eq:euler_ineq} reduces the left-hand side, so Assumption~\ref{ass:stability} implies the weaker condition with income included. We state the stronger form because it yields the linear lower bound for general utility (Proposition~\ref{prop:stab}) and is one component of \citet[Assumption~3.1]{ma2020income}, which is used for stationarity and tail behavior.
\end{remark}

\subsection{The Pareto trap}
\label{ssec:trap}

The positive linear lower bound in Assumption~\ref{a:1} follows from
the stability condition (Assumption~\ref{ass:stability}) alone, for any utility satisfying
Assumption~\ref{ass:existence}. Under CRRA utility, the same constant
$\bar{x}$ also enters the conditions for heavy-tailed stationary
wealth: the conditions that generate heavy tails simultaneously lock
in the lower bound.

\begin{proposition}[From stability to the linear lower bound]
\label{prop:stab}
If Assumptions~\ref{ass:existence} and~\ref{ass:stability} hold, then
the optimal consumption $f$ satisfies $f(w,s) \geq (1-\bar{x})w$ for
all $w > 0$ and $s \in \mathbb S$. In particular, Assumption~\ref{a:1}
holds with $a := 1 - \bar{x}$ and $w_0 = 0$.
\end{proposition}

\begin{proof}
This is \citet[Proposition~2.6]{ma2020income}. Assumption~\ref{ass:stability}
is their condition~(19), and their conclusion is the displayed lower
bound.
\end{proof}

By Theorem~\ref{thm:3}, whenever
stationary wealth and consumption have Pareto-type tails and the
stability condition holds, the ranking $\alpha_C > \alpha_W$ is
ruled out. The question is whether
stationary wealth can have a heavy tail in the first place, and
whether the conditions that produce it are compatible with the
stability condition.

\begin{definition}[Asymptotic MPC and saving rate]
\label{def:mpc}
Let $m_\infty(s) := \lim_{w \to \infty} f(w,s)/w$ whenever the limit
exists. Write $s_\infty(s) := 1 - m_\infty(s)$ for the asymptotic
saving rate. In the notation of Assumption~\ref{a:5},
$m_\infty(s) = K(s)$.
\end{definition}

\begin{remark}[The four roles of $\bar{x}$]\sloppy
\label{rem:roles_xbar}
The constant $\bar{x}$ from Assumption~\ref{ass:stability} enters the
\citet{ma2020income} framework at four levels:
\begin{enumerate}[label=(\roman*)]
  \item \emph{Linear lower bound.}
    Assumption~\ref{ass:stability} alone gives
    $f(w,s) \geq (1-\bar{x})w$ for any $u$ satisfying
    Assumption~\ref{ass:existence}
    (Proposition~\ref{prop:stab}).
  \item \emph{Existence of a stationary distribution.} Combined with
    a contraction-on-average condition on returns, stated formally
    in Assumption~\ref{ass:kesten}(i), the savings bound
    prevents wealth from drifting to infinity
    \citep[Theorem~3.1]{ma2020income}. This holds for any $u$
    satisfying Assumption~\ref{ass:existence}.
  \item \emph{Uniqueness and ergodicity.} With an additional mixing
    condition on income (Assumption~\ref{ass:kesten}(ii)), the
    stationary distribution is unique and ergodic
    \citep[Theorem~3.2]{ma2020income}, again for any $u$ satisfying
    Assumption~\ref{ass:existence}.
  \item \emph{Heavy-tailed wealth.} Under CRRA utility, the optimal
    saving function is asymptotically linear with rate $s_\infty(s)$
(Definition~\ref{def:mpc}),
    making wealth follow an approximately affine recursion at high
    levels. The classical theory of such recursions
    \citep{kesten1973random,goldie1991implicit} then supplies a
    Pareto-order lower bound for the wealth tail
    \citep[Theorem~3.3]{ma2020income}. Its expansion matrix uses the
    actual asymptotic saving rate $s_\infty(s)$, while $\bar{x}$ enters
    through the stationarity condition required by the theorem. This is the
    only level at which CRRA is used.
\end{enumerate}
\end{remark}

If the time-iteration (Coleman) operator
\citep[Section~2]{ma2020income}, which maps a candidate consumption
policy to the solution of the Euler equation, preserves concavity of
the policy function in wealth, then by
\citet[Proposition~2.5]{ma2020income} the optimal consumption $f$
is concave in $w$ for each $s$ and asymptotically linear
(Definition~\ref{df:asym_linear}(i)): $m_\infty(s)$ exists with
$m_\infty(s) \in [0,1]$. This concavity-preservation property holds
for CRRA utility, as shown in Remark~2.1 and Appendix~B of
\citet{ma2020income}.
More general asymptotic-linearity results for regularly varying marginal
utility, including HARA specifications, are available under additional
conditions \citep{ma2022asymptotic}. The stability condition
(Assumption~\ref{ass:stability}) sharpens the lower end: since
$f(w,s) \geq (1-\bar{x})w$ by Proposition~\ref{prop:stab}, concavity
gives $m_\infty(s) \geq 1 - \bar{x} > 0$.

Because $\mathbb S$ is finite, the convergence defining $m_\infty(s)$
is uniform across states. Hence
$(w-f(w,s))/w\to s_\infty(s)$ uniformly in $s$. Substitution into
\eqref{eq:budget} gives the high-wealth affine approximation
\begin{equation}\label{eq:kesten_recursion}
  W_{t+1} \approx A_{t+1} W_t + B_{t+1}, \qquad
  A_{t+1} := s_\infty(s_t)\, R_{t+1}, \quad B_{t+1} := Y_{t+1}.
\end{equation}
For the corresponding exact affine benchmark, define
$G_{ij}(\zeta):=s_\infty(s_i)R(s_j,\zeta)$ and the full-multiplier
Kesten matrix (cf.\ its calibration implementation in
Section~\ref{sec:calibration})
\begin{equation}
\label{eq:kesten_matrix}
  K_\alpha(i,j) := Q_{ij}\,\mathbb E\!\left[G_{ij}(\zeta)^\alpha\right],
  \qquad i,j = 1,\ldots,n.
\end{equation}
Under the classical conditions for an exact Markov affine recursion,
a positive solution to $r(K_\alpha)=1$, where $r(\cdot)$ denotes
spectral radius, characterizes its Pareto tail index
\citep{kesten1973random,goldie1991implicit,roitershtein2007one}.

The nonlinear income-fluctuation problem requires a distinct object.
Following \citet[Theorem~3.3]{ma2020income}, define the expansion-only
matrix
\begin{equation}
\label{eq:mst_matrix}
  K^+_\kappa(i,j)
  :=Q_{ij}\,\mathbb E\!\left[
    \bigl(G_{ij}(\zeta)\,\mathbbm{1}\{G_{ij}(\zeta)>1\}\bigr)^\kappa
  \right]
\end{equation}
and $\lambda_+(\kappa):=r(K^+_\kappa)$. Its crossing point is
\[
  \kappa_W:=\inf\{\kappa>0:\lambda_+(\kappa)>1\}.
\]
This crossing point governs the Pareto-order lower bound established by
their theorem.

The superscript distinguishes $K^+_\kappa$ from the full-multiplier
matrix $K_\alpha$ in~\eqref{eq:kesten_matrix}. Only the
expansion events $G_{ij}(\zeta)>1$ enter $K^+_\kappa$, and
$\kappa_W$ need not equal the exact-affine tail index associated with
$K_\alpha$.

The following conditions, adapted from
\citet[Assumptions~3.1--3.3 and Theorem~3.3]{ma2020income}, ensure
stationarity and give a Pareto-order lower bound when the stationary
wealth distribution has unbounded support:

\begin{assumption}[Conditions for the heavy-tail lower bound]%
\footnote{Under the notational correspondence
$\mathbb S = Z$, $Q = P$, $w = a$, $f = c^*$,
and our $\bar{x}$ is their $\bar{s}$, with \citet{ma2020income}:
Assumptions~\ref{ass:stability} and~\ref{ass:kesten}(i)
restate their Assumption~3.1,
condition~(ii) restates their Assumption~3.2, and
conditions~(iii)-(iv), together with the direct asymptotic-MPC
condition imposed below, provide the properties used in their
Assumption~3.3 and Theorem~3.3.}
\label{ass:kesten}
Let $\bar{x} \in (0,1)$ be the constant from
Assumption~\ref{ass:stability}.
\begin{enumerate}
  \item[(i)] \textbf{Contraction on average.}
    $\bar{x}\,\lim_{n\to\infty}(\EE\prod_{t=1}^{n} R_t)^{1/n} < 1$.
  \item[(ii)] \textbf{Income support.} There exists
    $\bar{s} \in \mathbb S$ with $Q_{\bar{s}\bar{s}} > 0$. Let
    $y_\ell \geq 0$ be the greatest lower bound of the support of
    $\{Y_t\}$. Either the support is finite and
    $\PP(Y_t = y_\ell \mid s_t = \bar{s}) > 0$, or $\{Y_t\}$ admits
    a conditional density that is positive on
    $(y_\ell, y_\ell + \delta)$ at $\bar{s}$ for some $\delta > 0$.
  \item[(iii)] \textbf{Spread.} For the same $\bar{s}$ as in~(ii),
    $\PP(s_\infty(\bar{s})\,R(\bar{s},\zeta) > 1) > 0$.
  \item[(iv)] \textbf{Crossing.} There exists $\theta > 0$ such that
    $1 < \lambda_+(\theta) < \infty$.
\end{enumerate}
\end{assumption}

Condition~(i) uses the \emph{same} $\bar{x}$ from the stability
condition (Assumption~\ref{ass:stability}): the savings upper bound,
combined with the long-run return, contracts wealth on average.
Condition~(ii) ensures mixing (social mobility): currently rich
households eventually receive low income and decline.
Condition~(iii) requires occasional multiplicative expansion: the
effective return $s_\infty(\bar{s})\,R$ exceeds one with positive
probability in at least one persistent exogenous state.
Condition~(iv) requires that this expansion pushes $\lambda_+(\kappa)$
above one at some finite $\kappa$.

If Assumptions~\ref{ass:existence}, \ref{ass:stability},
and~\ref{ass:kesten} hold and $m_\infty(s)\in(0,1)$ exists for every
$s\in\mathbb S$ (Definition~\ref{def:mpc}), then the arguments of
\citet[Theorems~3.1--3.3]{ma2020income} apply. Their Assumption~3.3
obtains the asymptotic MPC through Proposition~2.5. Here it is imposed
directly. A unique stationary wealth distribution exists, and if it
has unbounded support, then for every $\varepsilon > 0$,
\begin{equation}\label{eq:mst_tail}
  \liminf_{w \to \infty}\; w^{\kappa_W + \varepsilon}\,
  \PP(W > w) \;>\; 0.
\end{equation}
Combining the linear lower bound from Proposition~\ref{prop:stab}
with~\eqref{eq:mst_tail} yields the following model-specific
impossibility result.

\begin{theorem}[The Pareto trap]\label{thm:trap}
If Assumptions~\ref{ass:existence}, \ref{ass:stability},
and~\ref{ass:kesten} hold, $m_\infty(s)\in(0,1)$ exists for every
$s\in\mathbb S$ (Definition~\ref{def:mpc}), and
stationary wealth $W$ has unbounded support, then $C$ cannot have a
Pareto-type tail with index $\alpha_C > \kappa_W$.
\end{theorem}

\begin{proof}
By Proposition~\ref{prop:stab},
$C = f(W,S) \geq (1-\bar{x})W$ almost surely, where
$\bar{x} \in (0,1)$ is the constant from
Assumption~\ref{ass:stability}. Hence for every $c > 0$,
\[
  \bar F_C(c) \;\geq\;
  \bar F_W\!\left(\frac{c}{1-\bar{x}}\right).
\]
Suppose to the contrary that $C$ has a Pareto-type tail with index
$\alpha_C > \kappa_W$, so that
$\bar F_C(c)=c^{-\alpha_C}L_C(c)$ for a slowly varying function $L_C$.
By~\eqref{eq:mst_tail}, for every $\varepsilon > 0$,
\[
  \liminf_{c \to \infty}\; c^{\kappa_W + \varepsilon}\,
  \bar F_C(c)
  \;\geq\; (1-\bar{x})^{\kappa_W + \varepsilon}\,
  \liminf_{c \to \infty}\;
  \left(\frac{c}{1-\bar{x}}\right)^{\!\kappa_W + \varepsilon}\,
  \bar F_W\!\left(\frac{c}{1-\bar{x}}\right)
  \;>\; 0.
\]
Pick $\varepsilon > 0$ with $\kappa_W + \varepsilon < \alpha_C$.
Then
\[
  c^{\kappa_W + \varepsilon}\, \bar F_C(c) \;=\;
  c^{-(\alpha_C-\kappa_W-\varepsilon)}L_C(c).
\]
Because every positive power dominates a slowly varying function
\citep[Proposition~1.3.6]{bingham1987regular}, the right-hand side tends
to zero. Hence
\[
  \lim_{c \to \infty} c^{\kappa_W + \varepsilon}\bar F_C(c) = 0,
\]
contradicting the positive $\liminf$ above.
\end{proof}

\begin{corollary}[The Pareto trap under CRRA]\label{cor:trap_crra}
If $u$ is CRRA with $\gamma > 0$ and
Assumptions~\ref{ass:existence}, \ref{ass:stability},
and~\ref{ass:kesten} hold, then $m_\infty(s)\in(0,1)$ exists for
every $s\in\mathbb S$, and Theorem~\ref{thm:trap} applies whenever
stationary wealth has unbounded support.
\end{corollary}

\begin{proof}
CRRA satisfies the concavity-preservation property (condition~(17)
of \citealt{ma2020income}, their Remark~2.1 and Appendix~B). By
\citet[Proposition~2.5]{ma2020income}, $f$ is concave in $w$ and
$m_\infty(s) := \lim_{w \to \infty} f(w,s)/w$ exists with
$m_\infty(s) \in [0,1]$.

\emph{Lower bound.} Since $f(w,s) \geq (1-\bar{x})w$
(Proposition~\ref{prop:stab}) with $\bar{x} \in (0,1)$, we have
$f(w,s)/w \geq 1-\bar{x}$ for all $w > 0$. Taking $w \to \infty$
gives $m_\infty(s) \geq 1 - \bar{x} > 0$.

\emph{Upper bound.} Suppose $m_\infty(s_i) = 1$ for some
$i\in\{1,\ldots,n\}$. Since $f(\cdot, s_i)$ is concave with
$f(0, s_i) \geq 0$, the ratio $f(w, s_i)/w$ is nonincreasing in
$w$ on $(0, \infty)$. The ratio never exceeds one (feasibility) and has limit one, so it
equals one for all $w>0$. Hence $f(w, s_i) = w$ for all
$w > 0$, meaning the household chooses zero saving at state $s_i$.
The optimality condition
\citep[equation~(10)]{ma2020income} at the corner $c = w$ requires
\[
u'(w) \;\geq\; \sum_{j=1}^{n} Q_{ij}\,
\EE\!\left[\beta(s_j,\varepsilon)\, R(s_j,\zeta)\,
u'\!\bigl(f(Y(s_j,\eta), s_j)\bigr)\right].
\]
Since $f(Y(s_j,\eta), s_j) \leq Y(s_j,\eta)$ (feasibility) and
$u'' < 0$, the right-hand side is bounded below by
$\sum_{j} Q_{ij}\, \EE[\beta(s_j,\varepsilon)\,
R(s_j,\zeta)\, u'(Y(s_j,\eta))]$, which is strictly positive by
$\beta,R,u'>0$. Under CRRA,
$u'(w) = w^{-\gamma} \to 0$ as $w \to \infty$, so there exists a
finite wealth threshold beyond which the inequality fails. This contradicts optimality of $f$,
hence $m_\infty(s) < 1$ for all $s \in \mathbb S$.
\end{proof}

The next result establishes the linear lower bound for a
shifted-HARA case outside Assumption~\ref{ass:existence}(i).
Its finite marginal utility at zero requires the alternative policy
foundation of \citet{ma2022asymptotic}.

\begin{corollary}[Shifted HARA under an alternative foundation]\label{cor:trap_hara}
Suppose $u'(c)=(c+\bar c)^{-\gamma}$ with $\bar c>0$ and $\gamma>0$,
the processes $(\beta,R,Y)$ satisfy Assumption~2 of
\citet{ma2022asymptotic}, and Assumption~\ref{ass:stability}
holds for this $u$, with constant $\bar x$. Then the
income-fluctuation problem admits a unique optimal consumption
policy $f$, and $f(w,s)\geq(1-\bar x)w$ for every $w>0$ and
$s\in\mathbb S$, so Assumption~\ref{a:1} holds with
$a=1-\bar x$ and $w_0=0$. If, in addition, a stationary distribution for this
policy gives wealth and consumption Pareto-type tails, then
$\alpha_C\leq\alpha_W$.
\end{corollary}

\begin{proof}
Shifted HARA marginal utility is positive, continuous, strictly
decreasing, and vanishes at infinity, so Assumption~1 of
\citet{ma2022asymptotic} holds, and $u'(0)=\bar c^{-\gamma}$ is
finite and admissible in their framework (their Lemma~2.1). Within
their Assumption~2, the two marginal-utility moments are finite for
this $u$ because $u'\leq\bar c^{-\gamma}$ and the conditional
moments of $\beta R$ are finite, while the income moment
$\EE Y<\infty$ is maintained. Their Theorem~2.2(ii) delivers the
unique optimal policy $f$ as the limit in their marginal-utility
metric of $T^{n}c_0$ for any initial candidate $c_0$, where $T$ is
the policy-iteration operator of their Lemma~2.1.

The remaining argument specializes the proof of their Theorem~2.7
to the constant fraction $1-\bar{x}$ and threshold zero. Let
$\mathcal C^{*}$ be the set of candidate consumption functions $c$
with $c(w,s)\geq(1-\bar{x})w$ for every $w>0$ and $s\in\mathbb S$.
The initial candidate $c_0(w,s)=w$ belongs to $\mathcal C^{*}$. The
class is invariant under $T$. Fix $c\in\mathcal C^{*}$, $w>0$, and
$s_i\in\mathbb S$, and write $\xi=Tc(w,s_i)$. Suppose
$\xi<(1-\bar{x})w$. Then $w-\xi>\bar{x}w$, so every realized
next-period wealth satisfies
$R(s_j,\zeta)(w-\xi)+Y(s_j,\eta)\geq R(s_j,\zeta)\,\bar{x}\,w$,
and the bound defining $\mathcal C^{*}$ gives
$u'\bigl(c(R(s_j,\zeta)(w-\xi)+Y(s_j,\eta),s_j)\bigr)
\leq u'\bigl(R(s_j,\zeta)\,\bar{x}\,(1-\bar{x})\,w\bigr)$.
The Euler update of their Lemma~2.1 therefore yields
\[
  u'(\xi)\leq\max\Bigl\{\sum_{j=1}^{n} Q_{ij}\,
  \EE\bigl[\beta(s_j,\varepsilon)\,R(s_j,\zeta)\,
  u'\bigl(R(s_j,\zeta)\bar{x}(1-\bar{x})w\bigr)\bigr],\,
  u'(w)\Bigr\}
  \leq u'\bigl((1-\bar{x})w\bigr),
\]
where the sum is the left-hand side of
Assumption~\ref{ass:stability} evaluated at wealth
$(1-\bar{x})w$ and $u'(w)\leq u'((1-\bar{x})w)$ directly. Hence
$\xi\geq(1-\bar{x})w$, a contradiction, so
$Tc(w,s_i)\geq(1-\bar{x})w$. By induction
$T^{n}c_0\in\mathcal C^{*}$ for every $n$, convergence in the
marginal-utility metric implies pointwise convergence, and the bound
passes to the limit, so $f(w,s)\geq(1-\bar{x})w$. The tail
restriction follows from Theorem~\ref{thm:3} with
$a=1-\bar{x}$ and $w_0=0$.
\end{proof}

\begin{remark}[Scope of the shifted-HARA extension]\label{rem:hara_scope}
The corollary establishes the policy bound. Its tail-index
implication is conditional on a stationary distribution with the stated
Pareto-type tails. Neither existence of that distribution nor construction
of the wealth tail is established here. At $\bar c=0$ the utility
is CRRA and the stronger Corollary~\ref{cor:trap_crra}
applies.
\end{remark}

The optimal policy satisfies $0\leq f(w,s)\leq w$, so the stationary
pair satisfies $\PP\{C\leq W\}=1$: the model's global feasibility
bound verifies the almost-sure assumption of the next corollary.

\begin{corollary}[Feasibility and two-sided inheritance]
\label{cor:feas_inherit}
Suppose Assumption~\ref{a:1} holds and $\PP\{C\leq W\}=1$. If
$W$ has a Pareto-type tail with index $\alpha_W$, then $C$ has a weak
Pareto-type tail with index $\alpha_W$.
\end{corollary}

\begin{proof}
$\PP\{C\leq W\}=1$ gives eventual tail domination
(Lemma~\ref{lem:tail_domination}). Apply
Corollary~\ref{cor:weak_inheritance} with
Assumption~\ref{a:1}.
\end{proof}

Thus the weak tail index is inherited without independence.
Independence remains relevant to the stronger Pareto-tail conclusion
and tail constant in Proposition~\ref{th:2}.

\begin{remark}[Tail bounds vs.\ Pareto tails]
\label{rem:mst_vs_exact}
Equation~\eqref{eq:mst_tail} is a $\liminf$ bound, not an exact tail.
For an exact affine recursion, the
classical theory can deliver a Pareto tail under its additional
moment, irreducibility, and nonarithmetic conditions
\citep{kesten1973random,goldie1991implicit,buraczewski2016stochastic}.
The root of the full-multiplier matrix $K_\alpha$ is the corresponding
exact-affine tail index, while $\kappa_W$ comes from the expansion-only
matrix $K^+_\kappa$ used in the nonlinear lower-bound argument. If the
stationary wealth distribution itself has a Pareto-type index
$\alpha_W$, Theorem~\ref{thm:3} gives the sharper and directly
comparable restriction $\alpha_C\leq\alpha_W$.
\end{remark}

\begin{remark}[Scope]\label{rem:trap_scope}
Subject to its remaining assumptions, Theorem~\ref{thm:trap}
applies to any utility under which $m_\infty(s)\in(0,1)$ exists in
every state.
Corollary~\ref{cor:trap_crra} covers CRRA specifications
within the \citet{ma2020income} framework whenever the stated
assumptions hold, and Corollary~\ref{cor:trap_hara} provides
the shifted-HARA policy bound under the alternative foundation of
\citet{ma2022asymptotic}. Appendix~\ref{app:survey} classifies related
heterogeneous-agent models \citep{benhabib2011distribution,benhabib2015wealth,
benhabib2016distribution,toda2019wealth,hubmer2021sources,kaymak2016evolution}
against the theorem's conditions.
\end{remark}

\begin{definition}[Stationary recursive competitive equilibrium]
\label{def:srce}
Given primitives, a stationary recursive competitive equilibrium
consists of a price vector, a household policy $f$, and a
distribution $\phi$. At those prices, $f$ solves the household
problem, $\phi$ is invariant under the Markov kernel induced by $f$
and the exogenous state process, asset and goods markets clear, and
prices satisfy firm optimality.
\end{definition}

\begin{remark}[General equilibrium]
\label{rem:ge}
The impossibility is proved conditional on a price vector. In a
stationary recursive competitive equilibrium
(Definition~\ref{def:srce}), prices are selected so that the
household problem generates an invariant distribution consistent
with market clearing. At those prices, the household problem
is the income-fluctuation
problem~\eqref{eq:objective}-\eqref{eq:budget}, so
Theorem~\ref{thm:trap} applies whenever
Assumptions~\ref{ass:existence}, \ref{ass:stability},
and~\ref{ass:kesten} hold there, $m_\infty(s)\in(0,1)$ exists in
every state, and stationary wealth has unbounded support. The stability
condition is part of the sufficient conditions for stationarity in
\citet[Theorem~3.1]{ma2020income}, not a necessary condition for every
stationary equilibrium. General equilibrium can therefore affect
whether the hypotheses of the Pareto trap hold, but it does not alter
the conditional tail implication once they do.
\end{remark}

\subsection{A second application: finitely lived households}
\label{ssec:olg}

The income-fluctuation application uses the cash-on-hand bound
$C\leq W$. The Pareto trap does not depend on that feature. In the
finitely lived overlapping-generations economy of
\citet{benhabib2011distribution}, the optimal consumption rule at
age $\tau$ is $C=m(\tau,R)\{W+H(\tau,R,Y)\}$, where $H$ is
remaining human wealth \citep[Appendix~A]{benhabib2011distribution},
so consumption can exceed current financial wealth.

\begin{proposition}[OLG population tails with human wealth]
\label{prop:olg}
Impose Assumptions~1--3 and the regularity conditions of
Theorems~1--2 of \citet{benhabib2011distribution}. In their
stationary population cross-section, financial wealth satisfies
the two-sided power-tail bound
\[
  0<\liminf_{v\to\infty}v^{\mu}\bar F_W(v)
  \leq\limsup_{v\to\infty}v^{\mu}\bar F_W(v)<\infty
\]
for some $\mu>1$, and consumption satisfies the same two-sided
bound with the same exponent. Hence $W$ and $C$ have weak
Pareto-type tails with index $\mu$, and if $C$ has a Pareto tail,
its index is $\mu$.
\end{proposition}

\begin{proof}
Their Lemma~A.2 gives a unique ergodic dynasty process, their
Theorem~1 gives an exact Pareto tail for the initial-wealth law,
and their Theorem~2 transfers the power exponent $\mu$ to the
stationary population wealth distribution in the stated two-sided
form. For each of the finitely many return states, the propensity
$m(\tau,R)$ is positive and continuous on the compact age
interval, so across ages and return states it attains a positive
minimum $m_-$ and a finite maximum $m_+$, and with finitely many
income states remaining human wealth is bounded by some
$H_+<\infty$. The affine rule then gives
\[
  m_-W \;\leq\; C \;\leq\; m_+W+m_+H_+
  \qquad\text{almost surely}.
\]
For $x>0$, $\{W>x/m_-\}\subseteq\{C>x\}$, so
$\bar F_C(x)\geq\bar F_W(x/m_-)$, and for $x>m_+H_+$,
$\{C>x\}\subseteq\{W>(x-m_+H_+)/m_+\}$, so
$\bar F_C(x)\leq\bar F_W((x-m_+H_+)/m_+)$. Both rescalings
preserve two-sided $v^{\mu}$ bounds:
$x^{\mu}\bar F_W(x/m_-)=m_-^{\mu}(x/m_-)^{\mu}\bar F_W(x/m_-)$,
and $(x-m_+H_+)/m_+\sim x/m_+$ makes the upper comparison
asymptotically a fixed rescaling. The two-sided bound for $C$
follows. The two-sided bounds give the logarithmic limits, so both
weak Pareto-type indices equal $\mu$, and a Pareto tail for $C$
pins its index at its weak index.
\end{proof}

Thus the Pareto trap survives $C>W$ when the extra resource is
bounded, with the model's own ergodicity and power-tail theorems
supplying the stationary distribution. Their Theorem~1 concerns the
initial-wealth law and their Theorem~2 the stationary population
law, and their Claim~1 obtains its exact population Pareto
constant under the assumption of i.i.d.\ returns.

\subsection{Escape mechanisms}
\label{ssec:escape}

Subsection~\ref{ssec:escape_criterion} states the necessary escape
criterion: the positive linear lower bound must fail. This paper
states the criterion and does not develop escape mechanisms. One
published example satisfies it: under power wealth-in-utility
preferences with $\delta<\gamma$,
\citet[Theorem~2]{ma2025optimal} establish, under their additional
conditions, including irreducibility, a spectral-radius
restriction, an income floor, and positive probability of a
positive discounted return from every state, that
$f(w,s)$ grows like $w^{\delta/\gamma}$, which is uniform
asymptotic power growth with exponent $\rho=\delta/\gamma<1$ and,
for wealth with a weak Pareto-type index $\alpha_W$, maps to the
weak consumption index $\alpha_C=\alpha_W\gamma/\delta$ through
Theorem~\ref{thm:tail_mapping} under its local-boundedness
condition. When $\delta\geq\gamma$, the
asymptotic MPC is strictly positive under the respective conditions
of \citet[Theorems~3--4]{ma2025optimal}, and the trap logic applies
under the remaining hypotheses of Theorem~\ref{thm:trap}.
\citet[Chapter~5]{hu2026pareto} develops the wealth-motive and
liquidation-cost mechanisms. In a separate return-risk example,
\citet{ma2021theory} obtain a zero asymptotic MPC at sufficiently
high risk aversion. Appendix~\ref{app:survey} records the
open model classes without assigning them an escape status.
Section~\ref{sec:calibration} quantifies the trap in the calibrated
benchmark.

\section{Calibration: the trap bites}
\label{sec:calibration}

This section calibrates the benchmark of Section~\ref{sec:ha_models} to each eligible wealth-tail index, quantifies the resulting implications for consumption, and states which empirical comparisons permit each calculation. Every empirical input comes from the prespecified release of Section~\ref{sec:empirics}, and the calibration carries forward the empirical covariance matrix, multiplicity decision, and object-specific eligibility recorded there. Index-calibration eligibility does not permit a top-share calculation under a Pareto distribution.

\subsection{Calibration identities}
\label{ssec:calib_identities}

The calibration runs from the wealth index to the return mean, inverting the asymptotic theory of Section~\ref{sec:ha_models}: with lognormal returns, fixed preferences, and fixed return dispersion, the wealth target pins down the mean log return. Specialize to i.i.d.\ returns and income, a constant discount factor $\beta\in(0,1)$, and CRRA utility with $\gamma>0$, including log utility, $\gamma=1$. Maintain an interior asymptotic policy limit: as cash on hand grows, the fraction consumed approaches $m_\infty\in(0,1)$. The asymptotic saving rate then solves the high-wealth Euler identity
\begin{equation}
\label{eq:sinf_formula}
    1=\beta s_\infty^{-\gamma}\,\EE[R^{1-\gamma}],
    \qquad
    s_\infty=\bigl(\beta\,\EE[R^{1-\gamma}]\bigr)^{1/\gamma},
\end{equation}
obtained by substituting $c_t\approx m_\infty w_t$ and $c_{t+1}\approx m_\infty s_\infty R\,w_t$ into the Euler equation and canceling the common factor. The saving rate is the complementary share, $s_\infty=1-m_\infty$, and in the log case the identity gives $s_\infty=\beta$. For the corresponding exact affine recursion, the positive Kesten root solves
\begin{equation}
\label{eq:alpha_equation}
    \EE[A^{\alpha}]=\EE[(s_\infty R)^{\alpha}]=1,
\end{equation}
the scalar case of the spectral condition $r(K_{\alpha})=1$ in equation~\eqref{eq:kesten_matrix}. For lognormal returns, $\log R\sim N(\mu_R,\sigma_R^2)$ with $\mu_R$ the mean log return, distinct from the mean gross return $\EE[R]=\exp(\mu_R+\sigma_R^2/2)$, the two identities solve in closed form for the mean log return that places the analytical root at a target $\alpha_W$:
\begin{equation}
\label{eq:mu_lognormal}
    \mu_R
    =
    -\ln\beta
    -\frac{1}{2}(1-\gamma)^2\sigma_R^2
    -\frac{1}{2}\gamma\,\alpha_W\,\sigma_R^2 .
\end{equation}
Thus $(\beta,\gamma,\sigma_R,\alpha_W)$ pin down $\mu_R$ analytically. The identities and the verification protocol follow \citet[Chapter~2]{hu2026pareto}. Given each eligible wealth index, equation~\eqref{eq:mu_lognormal} therefore supplies the calibrated return mean directly. The calibration targets wealth alone, and the model's consumption tail is an untargeted prediction throughout.

\subsection{Setup and calibration strategy}
\label{ssec:calib_setup}

We calibrate the standard income-fluctuation problem of Section~\ref{sec:ha_models} to each eligible wealth-tail index and examine its untargeted implication for consumption. The empirical inputs are Australia (HILDA wave~22) and Germany, Italy, and the Netherlands (HFCS wave~5). Table~\ref{ch3:tab:scorecard} records their estimates and reporting status. The model restriction is asymptotic: under the maintained tail conditions, the positive linear lower bound rules out $\alpha_C>\alpha_W$, while asymptotic linearity and independence sharpen the prediction to $\alpha_C=\alpha_W$.

The benchmark has CRRA utility $u(c)=c^{1-\gamma}/(1-\gamma)$ and i.i.d.\ lognormal shocks,
\[
  \log R\sim N(\mu_R,\sigma_R^2),
  \qquad
  \log Y\sim N(\mu_Y,\sigma_Y^2).
\]
We set $\gamma=1.5$, $\beta=0.95$, $\sigma_R=0.20$, $\sigma_Y=0.50$, and normalize mean income to one. These are maintained illustrative values, not a joint structural estimate of preferences and both shock processes. For each eligible $\alpha_W$, the code calibrates $\mu_R$ using the closed form~\eqref{eq:mu_lognormal} and imposes the scalar Kesten condition~\eqref{eq:alpha_equation}, where $s_\infty$ is the analytical asymptotic saving rate of equation~\eqref{eq:sinf_formula}. It then solves the nonlinear problem by the endogenous grid method \citep{carroll2006method} on a 600-point log grid. The stopping rule is a scale-adjusted sup norm of $10^{-11}$. Write $s_{\rm term}=1-f(w_{\max}^{\rm policy})/w_{\max}^{\rm policy}$ for the saving rate at the last policy-grid point. Separate checks cover the maximum relative residual of the interior Euler equation~\eqref{eq:euler_policy} (at most $10^{-3}$), feasibility ($0<f(w)\le w+10^{-10}$), monotonicity (decreases of $f$ at most $10^{-10}$), contraction ($\EE[\log(s_{\rm term}R)]<0$), the Kesten moment evaluated with $s_{\rm term}$ at the target (within $10^{-3}$ of one), and the positive root implied by $s_{\rm term}$ (within $10^{-2}$ of $\alpha_W$).

\begin{table}[H]
\centering
\caption{Empirical targets and object-specific reporting status}
\label{ch3:tab:scorecard}
\begin{tabular}{lcccccc}
\toprule
Country & $\hat\alpha_W$ & $\hat\alpha_C$ & Gap (s.e.) & Holm $p$ & Reject & Exact tail \\
\midrule
Australia & 2.119 & 3.168 & 1.050 (0.199) & $<0.001$ & yes & no \\
Germany & 1.603 & 4.085 & 2.482 (0.276) & 0.002 & yes & yes \\
Italy & 1.615 & 3.383 & 1.768 (0.223) & $<0.001$ & yes & yes \\
Netherlands & 2.096 & 2.906 & 0.810 (0.271) & 0.081 & no & no \\
\bottomrule
\end{tabular}
\begin{flushleft}
\footnotesize
All rows meet the index-calibration requirements. The Holm decision concerns the one-sided gap in the fixed weighted-survey family and is distinct from index eligibility. ``Exact tail'' permits pure-Pareto exponent benchmarks only. It does not identify a population top share without below-threshold aggregation.
\end{flushleft}
\end{table}

\subsection{Policy diagnostics}
\label{ssec:calib_policy}

Table~\ref{ch3:tab:calibration} reports the calibrated mean log return and the policy diagnostics. The consumption and saving shares in the table come from the last policy-grid point and agree with the analytical limits at the displayed precision. The last column evaluates $\EE[(s_{\rm term}R)^{\alpha_W}]$ and rounds to one in all four cases. All four calibrations pass the prespecified numerical checks. Equation~\eqref{eq:sinf_formula} gives a limiting saving rate below one in each case. Proposition~\ref{prop:stab} supplies the global lower bound used by Theorem~\ref{thm:3}.

\begin{table}[H]
\centering
\caption{Calibrated standard income-fluctuation parameters and policy diagnostics}
\label{ch3:tab:calibration}
\begin{tabular}{lcccccc}
\toprule
Country & $\alpha_W$ & $\mu_R$ & $\EE[R]$ & $m_\infty$ & $s_\infty$ & $\lambda_K(\alpha_W)$ \\
\midrule
Australia & 2.119 & -0.017 & 1.003 & 0.025 & 0.975 & 1.000 \\
Germany & 1.603 & -0.002 & 1.018 & 0.030 & 0.970 & 1.000 \\
Italy & 1.615 & -0.002 & 1.018 & 0.030 & 0.970 & 1.000 \\
Netherlands & 2.096 & -0.017 & 1.003 & 0.025 & 0.975 & 1.000 \\
\bottomrule
\end{tabular}
\begin{flushleft}
\footnotesize
Targets are the four index-calibration-eligible units in the prespecified empirical release. Maintained parameters are $\gamma=1.5$, $\beta=0.95$, $\sigma_R=0.20$, and $\sigma_Y=0.50$. Eligibility is conditional on the empirical index extrapolation. It does not imply exact-tail eligibility.
\end{flushleft}
\end{table}

\begin{figure}[H]
  \centering
  \includegraphics[width=\textwidth]
    {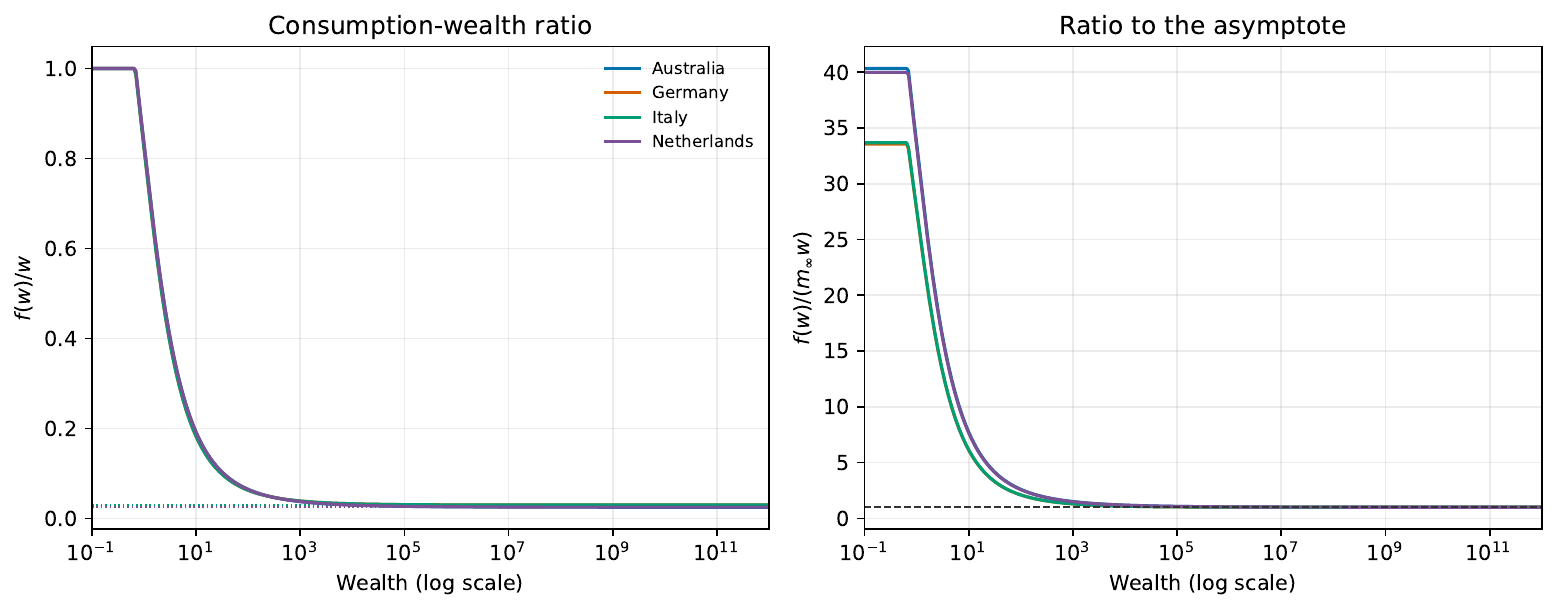}
  \caption[Policy diagnostics for the calibrations]{Policy diagnostics
  for the four index-calibration-eligible units. Dotted lines in the left panel
  mark the analytical asymptotic consumption-wealth ratios. The right
  panel shows the ratio to those limits.}
  \label{ch3:fig:policy}
\end{figure}

Figure~\ref{ch3:fig:policy} displays the solved policies. The left panel plots $f(w)/w$ and its analytical limit. The right panel normalizes by that limit. Convergence is slow: at one hundred times mean income the consumption share is still more than twice its asymptote in every calibration, and at ten thousand times mean income it remains about $9$ to $19$ percent above the limit. This distinction between a finite policy profile and its limiting slope is central to interpreting the simulations below.

\subsection{Exact-affine tail diagnostic}
\label{ssec:calib_sim}

To isolate the limiting multiplicative mechanism, we simulate the auxiliary recursion
\[
  W_{t+1}=s_{\rm term}R_{t+1}W_t+Y_{t+1},
\]
the numerical counterpart of the scalar high-wealth approximation in equation~\eqref{eq:kesten_recursion}, and evaluate the solved nonlinear consumption policy on the resulting wealth cross-section. This calculation uses $N=500{,}000$ agents, $T=8{,}000$ periods, a common seeded shock stream, and no empirical unit records. It is an exact-affine tail diagnostic, not the full nonlinear stationary economy.

\begin{figure}[H]
  \centering
  \includegraphics[width=0.72\textwidth]
    {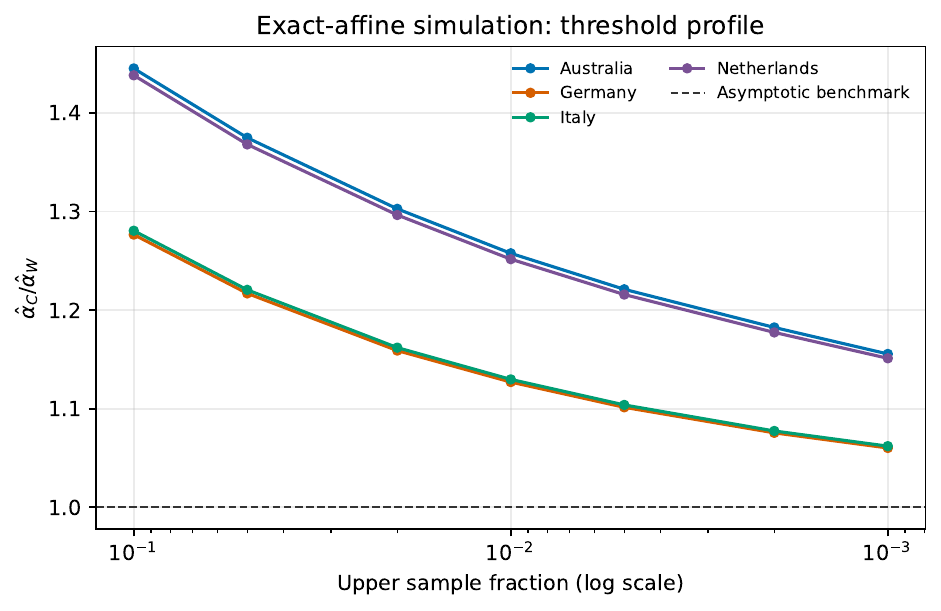}
  \caption[Exact-affine tail-index profiles]{Consumption-to-wealth Hill-index
  ratios in the exact-affine auxiliary simulations. The horizontal line is the
  asymptotic benchmark. Moving right uses a smaller fraction of the sample
  further into the tail.}
  \label{ch3:fig:tail_ratio_r2}
\end{figure}

Table~\ref{ch3:tab:sim_verification} reports fixed-fraction Hill estimates at the top 1\%. The simulated wealth index is close to its Kesten target, while the consumption-to-wealth index ratio remains above one at this finite threshold. Figure~\ref{ch3:fig:tail_ratio_r2} shows that the ratio moves toward one as the threshold advances further into the simulated tail. This pattern is consistent with exact asymptotic inheritance, but the finite simulation does not prove the limit.

\begin{table}[H]
\centering
\caption{Exact-affine simulation diagnostics at the 1\% tail fraction}
\label{ch3:tab:sim_verification}
\resizebox{\linewidth}{!}{
\begin{tabular}{lccccccc}
\toprule
Country & Target $\alpha_W$ & Sim.\ $\hat\alpha_W$ & Sim.\ $\hat\alpha_C$ & Ratio & Top 1\% $C$ & Gini($C$) & Var($\ln C$) \\
\midrule
Australia & 2.119 & 2.056 & 2.585 & 1.258 & 6.0\% & 0.266 & 0.184 \\
Germany & 1.603 & 1.559 & 1.757 & 1.127 & 13.0\% & 0.393 & 0.333 \\
Italy & 1.615 & 1.573 & 1.777 & 1.130 & 12.7\% & 0.389 & 0.327 \\
Netherlands & 2.096 & 2.032 & 2.544 & 1.252 & 6.2\% & 0.270 & 0.188 \\
\bottomrule
\end{tabular}}
\begin{flushleft}
\footnotesize
The exact-affine high-wealth approximation uses $N=500{,}000$ and $T=8{,}000$ with a common seeded shock stream. Hill estimates use a fixed 1\% sample fraction. Simulated top shares and body moments are model diagnostics, not empirical population shares. Ratios above one at this threshold are consistent with finite-range policy convergence and move toward one at deeper reported fractions. The full profile is shown in Figure~\ref{ch3:fig:tail_ratio_r2}.
\end{flushleft}
\end{table}

\subsection{Full nonlinear stationary distribution}
\label{ssec:calib_distribution}

\begin{figure}[H]
  \centering
  \includegraphics[width=\textwidth]
    {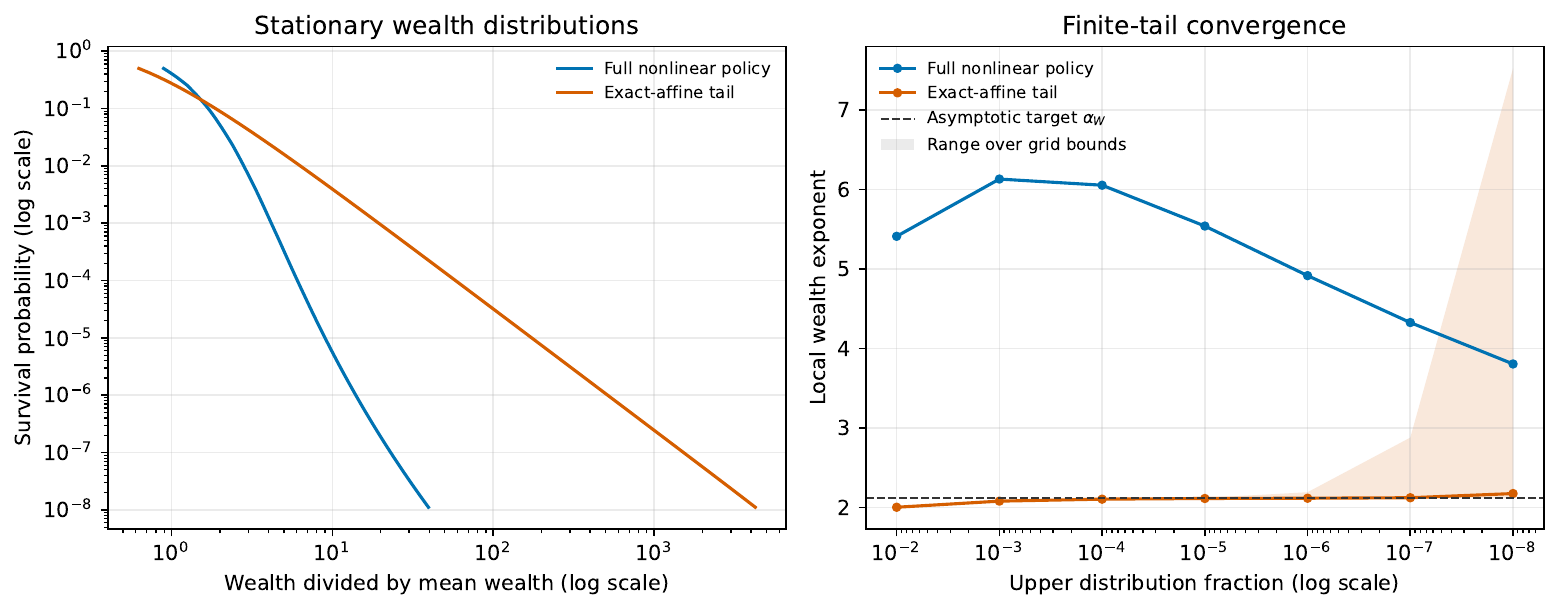}
  \caption[Full nonlinear and exact-affine stationary distributions]{Australian
  stationary-distribution diagnostics. The left panel plots survival probability against cash on hand divided by its mean for each process. The right panel plots the local cash-on-hand exponent against survival probability. The curves use the primary $10^6$ grid. Shading in the right panel spans the three grid ceilings, and the dashed line marks the calibrated Kesten target. The calculation describes behavior on these finite grids.}
  \label{ch3:fig:distribution_r2}
\end{figure}

The affine comparison applies the terminal saving rate at every wealth level as a numerical approximation to the limiting rule. To measure the resulting approximation error, we also iterate the stationary distribution under the full Australian nonlinear policy. The comparison uses the same shocks and quadrature rule for both processes. Table~\ref{ch3:tab:distribution_r2} and Figure~\ref{ch3:fig:distribution_r2} report the primary grid. Appendix~\ref{app:dist_iter} describes the calculation and repeats it over three upper wealth bounds.

\begin{table}[H]
\centering
\caption{Finite-grid distribution diagnostics for the Australian calibration}
\label{ch3:tab:distribution_r2}
\resizebox{\linewidth}{!}{
\begin{tabular}{lccccccc}
\toprule
Process & Mean $W$ & Gini($W$) & Top 1\% $W$ & Gini($C$) & Top 1\% $C$ & $\alpha_W(1\%)$ & $\alpha_W(0.01\%)$ \\
\midrule
Full nonlinear policy & 1.796 & 0.284 & 3.4\% & 0.116 & 1.6\% & 5.411 & 6.054 \\
Exact-affine tail & 45.194 & 0.475 & 12.3\% & 0.266 & 6.0\% & 2.004 & 2.106 \\
\bottomrule
\end{tabular}}
\begin{flushleft}
\footnotesize
The full process iterates the calibrated nonlinear policy. The exact-affine process imposes its limiting saving rate at every wealth level. The primary grid runs from $0.01$ to $1{,}000{,}000$. All entries are model diagnostics. Local exponents describe the indicated finite tail fractions and do not establish asymptotic convergence.
\end{flushleft}
\end{table}

Using $s_{\rm term}$ at every wealth level raises mean cash on hand from $1.8$ to $45.2$ times mean income and lowers the local wealth-tail index at the top 1\% from $5.41$ to $2.00$. Thus the exact-affine recursion changes body inequality and the onset of the tail even though it preserves the limiting Kesten mechanism. This is a finite-range finding, not a contradiction of the asymptotic theorem.

\subsection{The impossibility scorecard}
\label{ssec:calib_scorecard}

\begin{figure}[H]
  \centering
  \includegraphics[width=0.68\textwidth]
    {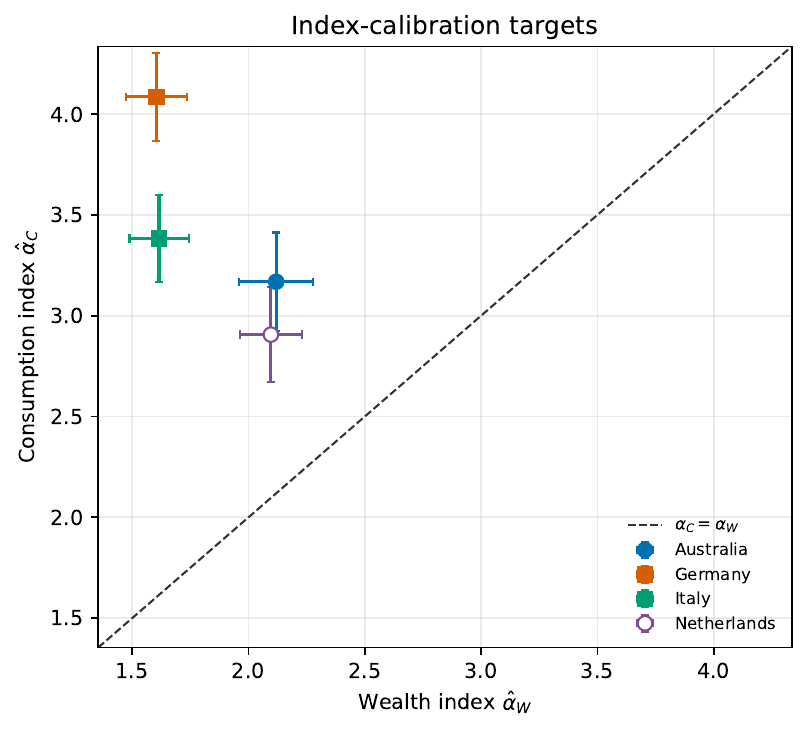}
  \caption[Empirical index scorecard]{Index-calibration targets. Filled
  markers reject the one-sided gap null after Holm adjustment. Square
  markers pass the separate exact-tail benchmark. Error bars are marginal standard errors
  and do not replace the joint gap inference.}
  \label{ch3:fig:scorecard_r2}
\end{figure}

Figure~\ref{ch3:fig:scorecard_r2} compares the four eligible index pairs with the i.i.d.\ asymptotic restriction $\alpha_C=\alpha_W$. All point estimates lie above the diagonal. In the fixed weighted-survey family, the one-sided Holm tests reject a nonpositive fitted consumption-minus-wealth gap for Australia, Germany, and Italy. The Dutch test does not reject. Under the maintained extrapolation to the limiting indices, the three rejections provide evidence against the benchmark restriction. Only Germany and Italy pass the separate screen for pure-Pareto benchmarks. An empirical population top share would also require the contribution below the fitted threshold.

\subsection{Robustness and scope}
\label{ssec:calib_robust}

\paragraph{Return volatility.}
Table~\ref{ch3:tab:return_volatility_r2} holds the Australian wealth-index target, preferences, and income process fixed while varying $\sigma_R$. The return mean is recalibrated in every row and the same numerical checks are applied. Greater volatility requires a lower mean return and a smaller asymptotic consumption share to preserve the target tail index. This is a controlled sensitivity, not a structural estimate of the return process.

\begin{table}[H]
\centering
\caption{Australian return-volatility sensitivity}
\label{ch3:tab:return_volatility_r2}
\begin{tabular}{lcccccc}
\toprule
$\sigma_R$ & $\mu_R$ & $\EE[R]$ & $m_\infty$ & $s_\infty$ & $\lambda_K(\alpha_W)$ & Euler residual \\
\midrule
0.15 & 0.013 & 1.024 & 0.036 & 0.964 & 1.000 & 0.00011 \\
0.20 & -0.017 & 1.003 & 0.025 & 0.975 & 1.000 & 0.00013 \\
0.25 & -0.056 & 0.976 & 0.010 & 0.990 & 0.999 & 0.00011 \\
\bottomrule
\end{tabular}
\begin{flushleft}
\footnotesize
Each row holds the wealth-index target, preferences, and income process fixed and recalibrates only the log-return mean. The extended policy grid is used for every row and all prespecified numerical checks pass. This controlled sensitivity is not a structural estimate of the return process.
\end{flushleft}
\end{table}

{\sloppy
\paragraph{Markov dependence.}
Theorem~\ref{thm:3} permits arbitrary dependence between wealth and the finite Markov state and gives the one-sided restriction whenever the existence, stability, lower-bound, and tail assumptions hold. Exact inheritance requires the stronger conditions stated in Proposition~\ref{th:2} and is not asserted for a generic Markov specification. No finite Markov calibration is reported because the transition parameters are not identified from the tail estimates.\par}

\paragraph{General equilibrium.}
The calibration is partial equilibrium. In a stationary recursive competitive equilibrium, prices are endogenous but fixed at their equilibrium values. The theorem applies at such a price vector only if its household-level assumptions hold there. General equilibrium can change both the validity of the stability condition and the numerical tail index. The condition must therefore be checked at the equilibrium prices, not assumed.

\subsection{Implications for model design}
\label{ssec:calib_implications}

The argument has four levels. First, Section~\ref{sec:empirics} identifies finite-window index pairs and their uncertainty. Second, the calibration interprets eligible pairs as asymptotic targets under a maintained tail extrapolation. Third, the theorem rules out those targets in the positive-lower-bound class when its assumptions hold. Finally, exact-tail top-share calculations require the additional exact-tail benchmark. Keeping these levels separate prevents finite-window evidence from being reported as an assumption-free asymptotic rejection.

For a uniformly locally bounded policy with uniform asymptotic power growth as in Definition~\ref{def:policy_growth}, Theorem~\ref{thm:tail_mapping} gives $\rho=\alpha_W/\alpha_C$ for the weak Pareto-type indices when wealth has such a tail. Under the maintained extrapolation, the released estimates of this ratio range from \ChThreeDeRho{} for Germany to \ChThreeNlRho{} for the Netherlands. They serve as targets for the escape mechanisms in \citet[Chapter~5]{hu2026pareto}, conditional on each mechanism's own eligibility requirements.

\section{Conclusion}
\label{sec:recap}

This paper treats the gap between consumption and wealth tails as a
specification restriction for heterogeneous-agent models.  The fitted
consumption index exceeds the fitted wealth index in all seven weighted-survey
comparisons of Section~\ref{sec:empirics}, consistent with existing U.S.\
evidence, and the difference survives the prespecified Holm adjustment for
Australia, Germany, and Italy.  The theoretical restriction does not depend on
those particular estimates.

The main result follows from a single property of the consumption policy.  If
consumption remains above a positive fraction of resources at the top, the
consumption tail cannot be strictly thinner than the resource tail.  No upper
feasibility bound or independence is needed for this one-sided impossibility.
Feasibility sharpens the conclusion to weak tail-index inheritance without
independence whenever wealth has a Pareto-type tail.  Separately, with an
exact Pareto wealth tail, independence and the affine or concave
asymptotically linear policy conditions of Proposition~\ref{th:2} yield a
Pareto consumption tail with an explicit constant.  These sharpenings answer different questions and neither narrows
the main theorem to a particular household budget.

The income-fluctuation application identifies the economic engine.  The
stability condition of \citet{ma2020income} supplies the difficult lower
bound, while cash-on-hand feasibility supplies the upper bound.  The same
stability certificate also enters their stationarity and heavy-tail argument,
creating the Pareto trap.  In the finitely lived OLG economy of \citet{benhabib2011distribution},
optimal consumption is affine in wealth, the same lower bound applies, and
the model's own population power-tail bounds transfer to consumption
(Proposition~\ref{prop:olg}).
Multi-asset and HANK models remain open here unless their lower exposure is
established rather than inferred by analogy.

The full-Pareto example in the introduction illustrates the possible scale: replacing an index of
$3.0$ by $1.5$ raises the analytical top 1\% benchmark from $4.6\%$ to
$21.5\%$.  The broader lesson is the overidentifying restriction.  Once a
model is calibrated to the wealth tail, its consumption tail is an untargeted
validation outcome.  A model that matches wealth but violates the consumption
ranking has not matched the joint inequality facts it was asked to explain.

\section*{Declaration of generative AI and AI-assisted technologies in the writing process}

During the late-stage preparation of this paper I used generative AI tools, principally Anthropic's Claude and OpenAI's Codex, working under my instruction at every step. The research questions, the theoretical results, and most of the manuscript predate my adoption of these tools. The tools polished the writing, checked statements, proofs, and cross-references, and helped debug the replication code I wrote. I reviewed and checked every output line by line, independently verified all mathematical results, and take full responsibility for the content of this paper.

\bibliographystyle{apalike}
\bibliography{reference}

\clearpage
\appendix

\section{Tail mapping results and proofs}
\label{app:imp_eta_proof}

This appendix collects the general mapping results used in Sections~\ref{sec:imp_key} and~\ref{sec:ha_models}: the asymptotic growth classes, the power-transformation and sandwich lemmas, the weak tail-mapping theorem, the Pareto tail constant, and the tail-domination results behind two-sided inheritance. It closes with the uniformity details for Proposition~\ref{th:2}. Throughout, $W$, $S$, and $C=f(W,S)$ are as in Section~\ref{sec:imp_key}.

\subsection*{Growth classes and regularity}

\begin{definition}[Asymptotic growth classes]
\label{df:asym_linear}
\label{def:policy_growth}
Let $g:\RR_+\to\RR_+$ be measurable. All limits in
parts~(i)-(iv) are taken as $x\to\infty$.
\begin{enumerate}[label=(\roman*)]
    \item $g$ is \emph{asymptotically linear} if
    \[
        \frac{g(x)}{x}\to k
        \qquad \text{for some } k>0 .
    \]

    \item $g$ is \emph{asymptotically sublinear} if
    \[
        \frac{g(x)}{x}\to 0 .
    \]

    \item $g$ has \emph{asymptotic power growth} with exponent
    $\rho>0$ if
    \[
        g(x)\sim hx^\rho
        \qquad \text{for some scalar } h>0 .
    \]

    \item A measurable function
    $f:\RR_+\times\mathbb S\to\RR_+$ has
    \emph{uniform asymptotic power growth} with exponent $\rho>0$ if
    there is a function $H:\mathbb S\to(0,\infty)$ such that
    \[
        \sup_{s\in\mathbb S}
        \left|
        \frac{f(w,s)}{w^\rho}-H(s)
        \right|
        \to 0
        \qquad \text{as } w\to\infty,
    \]
    where
    \[
        0<\inf_{s\in\mathbb S}H(s)
        \leq \sup_{s\in\mathbb S}H(s)<\infty.
    \]
\end{enumerate}
\end{definition}

For finite $\mathbb S$, the displayed bounds in part~(iv) follow automatically from $H(s)\in(0,\infty)$ for every $s$. Power growth with $\rho=1$ is equivalent to asymptotic linearity, and $0<\rho<1$ implies asymptotic sublinearity. The converse fails. For example, $g(x)=x/\log x$ is sublinear for large $x$ but has no asymptotic power growth with a positive finite coefficient, and $x^{1/2}$ and $x^{3/4}$ are both sublinear but encode different exponents. Sublinearity therefore supplies no unique tail-index mapping.

\begin{definition}[Uniform local boundedness]
\label{def:unif_loc_bounded}
A function $f:\RR_+\times\mathbb S\to\RR_+$ is \emph{locally bounded
in the first argument, uniformly in the second}, if
$\sup_{0\leq w\leq b,\,s\in\mathbb S}f(w,s)<\infty$ for every $b>0$.
\end{definition}

For brevity, we call such $f$ \emph{uniformly locally bounded}. Uniformity in $s$ is the only addition to the standard local-boundedness notion of regular variation theory \citep[cf.\ Corollary~1.4.2]{bingham1987regular}.

\subsection*{Power transforms and the weak tail mapping}

\begin{lemma}[Power transformation of a Pareto-type tail]
\label{lem:power_transform}
Suppose $h>0$ and $\rho>0$ are constants, $X\geq 0$, and $Y=hX^\rho$.
If $X$ has a Pareto-type tail with index $\alpha>0$, then $Y$ has a
Pareto-type tail with index $\alpha/\rho$. If, in addition, $X$ has a
Pareto tail with constant $k_X$, then $Y$ has a Pareto tail with
constant $k_X h^{\alpha/\rho}$. If instead $X$ has a weak
Pareto-type tail with index $\alpha$, then $Y$ has a weak Pareto-type
tail with index $\alpha/\rho$.
\end{lemma}

\begin{proof}
If $\bar F_X(x)=x^{-\alpha}L(x)$ with $L$ slowly varying, then
\[
        \PP\{hX^\rho>x\}
        =\bar F_X\left((x/h)^{1/\rho}\right)
        =h^{\alpha/\rho}x^{-\alpha/\rho}
          L\left((x/h)^{1/\rho}\right).
\]
The inner map $x\mapsto(x/h)^{1/\rho}$ is regularly varying with
index $1/\rho$ and tends to infinity. Its composition with $L$ is
therefore slowly varying
\citep[Proposition~1.5.7(ii)]{bingham1987regular}, so $Y$ is
Pareto-type with index $\alpha/\rho$. If $x^\alpha\bar F_X(x)\to k_X$,
the same calculation gives the stated Pareto-tail constant.
For the weak clause, put $u_x:=(x/h)^{1/\rho}$.  Then
$-\log\bar F_Y(x)/\log x
=(-\log\bar F_X(u_x)/\log u_x)(\log u_x/\log x)
\to\alpha/\rho$, since $\log u_x/\log x\to1/\rho$.
\end{proof}

\begin{lemma}[Sandwiching yields a weak Pareto-type tail]
\label{lem:sandwich}
Let $X,Y,Z$ be nonnegative random variables, and suppose $Y$ and $Z$
have Pareto-type tails with the same index $\tau>0$. If there exists
$x_0>0$ such that
\[
        \PP\{Y>x\}\leq \PP\{X>x\}\leq \PP\{Z>x\}
\]
for all $x\geq x_0$, then $X$ has a weak Pareto-type tail with index
$\tau$. In particular, the conclusion holds if $Y\leq X\leq Z$ almost
surely. The conclusion also holds if $Y$ and $Z$ have only weak
Pareto-type tails with index $\tau$.
\end{lemma}

\begin{proof}
Pareto-type tails satisfy the logarithmic limit, so
$-\log\bar F_Y(x)/\log x\to\tau$ and likewise for $Z$. The
survival ordering squeezes $-\log\bar F_X(x)/\log x$
between them. If $Y$ and $Z$ have only weak Pareto-type
tails, the logarithmic limits hold by definition and the
same squeeze applies.
\end{proof}

\begin{theorem}[Uniform asymptotic power growth and weak tail mapping]
\label{thm:tail_mapping}
Suppose $f$ is uniformly locally bounded
(Definition~\ref{def:unif_loc_bounded}) and has uniform
asymptotic power growth with exponent $\rho>0$ and limit function
$H:\mathbb S\to(0,\infty)$ in the sense of
Definition~\ref{def:policy_growth}(iv). If $W$ has a weak Pareto-type
tail with index $\alpha_W>0$, then $C$ has a weak Pareto-type tail
with index $\alpha_W/\rho$. Writing this weak index as $\alpha_C$
gives
\begin{equation}
\label{eq:alpha_mapping}
        \alpha_C=\frac{\alpha_W}{\rho}.
\end{equation}
\end{theorem}

\begin{proof}
Let $H_{\min}:=\inf_{s\in\mathbb S}H(s)$ and
$H_{\max}:=\sup_{s\in\mathbb S}H(s)$.
Definition~\ref{def:policy_growth}(iv) gives
$0<H_{\min}\leq H_{\max}<\infty$.
Fix $\varepsilon\in(0,H_{\min}/2)$. By uniform convergence, there is
$w_0>0$ such that, for all $w\geq w_0$ and all $s$,
\[
        (H(s)-\varepsilon)w^\rho
        \leq f(w,s)\leq
        (H(s)+\varepsilon)w^\rho .
\]
Hence, for all $w\geq w_0$ and all $s$,
\[
        \frac{H_{\min}}{2}w^\rho
        \leq f(w,s)\leq
        2H_{\max}w^\rho .
\]
Local boundedness gives
$\bar f_{w_0}:=\sup_{w\leq w_0,s\in\mathbb S}f(w,s)<\infty$.
Let $Y=(H_{\min}/2)W^\rho$ and $Z=2H_{\max}W^\rho$. Set
$x_0:=\max\{\bar f_{w_0},(H_{\min}/2)w_0^\rho\}$.
For every $x>x_0$, the event
$\{Y>x\}$ implies $W>w_0$, so the lower policy bound gives
$C\geq Y$. The event $\{C>x\}$ also implies $W>w_0$ because
$C\leq \bar f_{w_0}$ when $W\leq w_0$. The upper policy bound then gives
$C\leq Z$. Hence
\[
        \PP\{Y>x\}\leq \PP\{C>x\}\leq \PP\{Z>x\}
        \qquad \text{for all } x>x_0 .
\]
By the weak clause of
Lemma~\ref{lem:power_transform}, both $Y$ and $Z$ have
weak Pareto-type tails with index $\alpha_W/\rho$, and
Lemma~\ref{lem:sandwich} completes the proof.
\end{proof}

The theorem permits arbitrary dependence between $W$ and $S$, which matters in the economic applications, where stationary wealth and the idiosyncratic state are generally dependent. The price of that generality is a weaker conclusion: $C$ is only weak Pareto-type, though the theorem likewise needs only a weak index for $W$. The theorem gives a policy-level weak-tail formulation related to the exact index mappings of \citet[Propositions~1 and~3]{gaillard2023consumption} and to the policy limit of \citet[Theorem~2]{ma2025optimal}, whose quantitative section applies that limit to a calibrated wealth distribution.

\subsection*{Pareto tail constants}

Independent multiplicative factors are handled by Breiman's lemma
\citep{breiman1965some,resnick2007heavy}. If $X$ has a Pareto-type
tail with index $\alpha_X$, $M$ is a positive random variable
independent of $X$, and
$\EE[M^{\alpha_X+\varepsilon}]<\infty$
for some $\varepsilon>0$, then
$\bar F_{MX}(x)\sim\EE[M^{\alpha_X}]\bar F_X(x)$.
Multiplication by $M$ therefore scales the tail by $\EE[M^{\alpha_X}]$ and
leaves the Pareto-type index unchanged. The next result applies
this product form.

\begin{corollary}[Pareto tail constant]
\label{cor:exact_tail_mapping}
Suppose the conditions of Theorem~\ref{thm:tail_mapping} hold.
If, in addition, $W$ has a Pareto tail with index $\alpha_W$ and
constant $k_W$ and $W$ is
independent of $H(S)$, then $C=f(W,S)$ has a Pareto tail with index
$\alpha_W/\rho$ and constant
\begin{equation}
\label{eq:exact_tail_constant}
        k_C
        =
        k_W\,\EE\!\left[H(S)^{\alpha_W/\rho}\right].
\end{equation}
\end{corollary}

\begin{proof}
Set $\mathcal G:=H(S)W^\rho$, and let $H_{\min}$ and $H_{\max}$ be as
in the proof of Theorem~\ref{thm:tail_mapping}. For each
$s\in\mathbb S$, uniform convergence gives
$H(s)=\lim_{n\to\infty}f(n,s)/n^\rho$ along integer $n$. Hence $H$ is
measurable as a pointwise limit of the measurable sections
$s\mapsto f(n,s)/n^\rho$. Since $W$ is independent of $H(S)$ and
$W^\rho$ is a measurable function of $W$, the variables $W^\rho$ and
$H(S)$ are independent. By
Lemma~\ref{lem:power_transform}, $W^\rho$ has a Pareto tail with
index $\alpha_W/\rho$ and constant $k_W$. Moreover, $H(S)$ is
positive and bounded. Breiman's lemma, stated above, therefore gives
\[
        x^{\alpha_W/\rho}\PP\{\mathcal G>x\}
        \to
        k_{\mathcal G}
        :=k_W\EE\!\left[H(S)^{\alpha_W/\rho}\right].
\]
Fix $\delta\in(0,1)$. By uniform convergence and $H_{\min}>0$, there
is $w_\delta>0$ such that
\[
        \left|
        \frac{f(w,s)}{w^\rho}-H(s)
        \right|
        \leq \delta H_{\min}
        \leq \delta H(s)
        \qquad \text{for all } w\geq w_\delta \text{ and all } s,
\]
and hence
\[
        (1-\delta)H(s)w^\rho
        \leq f(w,s)\leq
        (1+\delta)H(s)w^\rho
        \qquad \text{for all } w\geq w_\delta \text{ and all } s.
\]
Set $\bar f_{w_\delta}:=\sup_{w\leq w_\delta,s\in\mathbb S}f(w,s)$
and
$x_\delta:=\max\{\bar f_{w_\delta},(1-\delta)H_{\max}w_\delta^\rho\}$.
Local boundedness gives $\bar f_{w_\delta}<\infty$. For every $x\geq x_\delta$, the
event $\{(1-\delta)\mathcal G>x\}$ implies
$H(S)W^\rho>H_{\max}w_\delta^\rho$, and since $H(S)\leq H_{\max}$
this forces $W>w_\delta$. The lower policy bound then gives
$C\geq(1-\delta)\mathcal G>x$. The event $\{C>x\}$ also implies
$W>w_\delta$ because $C\leq \bar f_{w_\delta}$ when
$W\leq w_\delta$. The
upper policy bound then gives $C\leq(1+\delta)\mathcal G$. Hence
\[
        \PP\{(1-\delta)\mathcal G>x\}
        \leq \bar F_C(x)
        \leq \PP\{(1+\delta)\mathcal G>x\}
        \qquad \text{for all } x\geq x_\delta .
\]
For every constant $\chi>0$,
\[
        x^{\alpha_W/\rho}\PP\{\chi\mathcal G>x\}
        =\chi^{\alpha_W/\rho}(x/\chi)^{\alpha_W/\rho}
          \PP\{\mathcal G>x/\chi\}
        \to \chi^{\alpha_W/\rho}k_{\mathcal G}.
\]
Applying this with $\chi=1-\delta$ and $\chi=1+\delta$ gives
\[
        (1-\delta)^{\alpha_W/\rho}k_{\mathcal G}
        \leq \liminf_{x\to\infty}x^{\alpha_W/\rho}\bar F_C(x)
        \leq \limsup_{x\to\infty}x^{\alpha_W/\rho}\bar F_C(x)
        \leq (1+\delta)^{\alpha_W/\rho}k_{\mathcal G}.
\]
Letting $\delta$ decrease to zero proves
\eqref{eq:exact_tail_constant}.
\end{proof}

Two sufficient conditions cover the applications. If $W$ is
independent of $S$, then $W$ is independent of $H(S)$, because
$H(S)$ is a measurable function of $S$. If the limit function is
constant, $H(s)=m>0$ for all $s$, independence holds automatically
and the constant is $k_C=k_Wm^{\alpha_W/\rho}$, with no restriction
on the dependence between $W$ and $S$.

\subsection*{Linear bounds and tail domination}

A one-sided upper bound gives complementary conclusions under less
structure: beyond a threshold, consumption does not exceed wealth,
the budget feasibility $c\leq w$.

\begin{condition}[Linear upper bound]
\label{cond:linear_upper_bound}
There exists $w_1\geq0$ such that
\[
        f(w,s)\leq w
        \qquad \text{for all } w\geq w_1
        \text{ and all } s\in\mathbb S .
\]
\end{condition}

When $w$ denotes resources available before consumption,
Condition~\ref{cond:linear_upper_bound} is the budget bound. The
witness $w_1=0$ is the global case, $f(w,s)\leq w$ for all $w\geq0$,
which forces $f(0,s)=0$, implies uniform local boundedness, and
implies $\PP\{C\leq W\}=1$. Models with another timing or resource
convention must verify the relevant bound directly.

\begin{definition}[Eventual tail domination]
\label{def:tail_dom}
$W$ \emph{eventually tail-dominates} $C$ if there exists
$x_0\geq0$ such that $\bar F_C(x)\leq\bar F_W(x)$ for every
$x\geq x_0$.
\end{definition}

For $x_0=0$, $C$ is smaller than $W$ in the usual stochastic order
\citep[cf.][Section~1.A.1, eq.~(1.A.1)]{shaked2007stochastic}.
The eventual form is what the proofs use.

\begin{lemma}[Two sufficient conditions for tail domination]
\label{lem:tail_domination}
\sloppy
\begin{enumerate}[label=(\roman*)]
    \item If Condition~\ref{cond:linear_upper_bound} holds and
    $f$ is uniformly locally bounded, then $W$ eventually
    tail-dominates $C$.
    \item If $\PP\{C\leq W\}=1$, then $W$ eventually tail-dominates
    $C$ with $x_0=0$.
\end{enumerate}
\end{lemma}

\begin{proof}
(i) Choose $b>w_1$ and set
$\bar f_b:=\sup_{0\leq w\leq b,\,s\in\mathbb S}f(w,s)$, which is
finite by Definition~\ref{def:unif_loc_bounded}. For
$x>\max\{\bar f_b,b\}$, the event $\{C>x\}$ implies $W>b$: otherwise
$C\leq \bar f_b<x$. On $\{W>b\}$,
Condition~\ref{cond:linear_upper_bound} gives $C\leq W$, since
$b\geq w_1$, so $W\geq C>x$. Hence
$\bar F_C(x)\leq\bar F_W(x)$ for every $x>\max\{\bar f_b,b\}$.
(ii) If $\PP\{C\leq W\}=1$, then $\{C>x\}\subseteq\{W>x\}$ up to a
null event for every $x\geq0$, so $\bar F_C(x)\leq\bar F_W(x)$ for
every $x\geq0$.
\end{proof}

\begin{corollary}[Two-sided bounds give weak tail-index inheritance]
\label{cor:weak_inheritance}
Suppose $W$ eventually tail-dominates $C$ and
Assumption~\ref{a:1} holds. If $W$ has a
weak Pareto-type tail with index $\alpha_W$, then $C$ has a weak
Pareto-type tail with index $\alpha_W$.
\end{corollary}

\begin{proof}
As in the proof of Theorem~\ref{thm:3},
$\bar F_W(x/a)\leq\bar F_C(x)$ for $x\geq aw_0$, and
domination gives $\bar F_C(x)\leq\bar F_W(x)$ for
$x\geq x_0$. Both outer logarithmic ratios tend to
$\alpha_W$, and the squeeze gives the limit for $C$.
\end{proof}

\subsection*{Uniformity details for Proposition~\ref{th:2}}

Proposition~\ref{th:2} follows from
Corollary~\ref{cor:exact_tail_mapping} once the policy is shown to
have uniform asymptotic power growth with exponent one. The remainder
of this appendix verifies that condition under each of the
proposition's two policy assumptions.

\paragraph{Exact affine policies.}
Under Assumption~\ref{a:3},
\[
    \frac{f(w,s)}{w}=h(s)+\frac{g(s)}{w}.
\]
Continuity on compact $\mathbb S$ makes $g$ bounded and gives constants
$0<h_{\min}\leq h(s)\leq h_{\max}<\infty$. Hence
\[
    \sup_{s\in\mathbb S}
    \left|\frac{f(w,s)}{w}-h(s)\right|
    \leq \frac{\sup_{s\in\mathbb S}g(s)}{w}
    \longrightarrow 0.
\]
The policy is locally bounded because, for each $b<\infty$,
\[
    \sup_{0\leq w\leq b,\,s\in\mathbb S}f(w,s)
    \leq h_{\max}b+\sup_{s\in\mathbb S}g(s)<\infty.
\]
Thus the conditions of Corollary~\ref{cor:exact_tail_mapping} hold
with $\rho=1$ and $H=h$.

\paragraph{Concave asymptotically linear policies.}
Under Assumption~\ref{a:5}, define
$r_w(s):=f(w,s)/w$ for $w>0$. Concavity of $f(\cdot,s)$ and
$f(0,s)\geq0$ imply that $r_w(s)$ is nonincreasing in $w$ for every
$s$. The sequence of continuous functions $r_n$ therefore decreases
pointwise on compact $\mathbb S$ to the continuous function $K$.
Dini's theorem \citep[Theorem~7.13]{rudin1976principles} gives
\[
    \sup_{s\in\mathbb S}|r_n(s)-K(s)|\longrightarrow0.
\]
For any $w\geq1$, let $n=\lfloor w\rfloor$. Since
$r_n(s)\geq r_w(s)\geq K(s)$,
\[
    0\leq
    \sup_{s\in\mathbb S}\{r_w(s)-K(s)\}
    \leq
    \sup_{s\in\mathbb S}\{r_n(s)-K(s)\}
    \longrightarrow0.
\]
Thus $f(w,s)/w\to K(s)$ uniformly in $s$. Continuity and positivity of
$K$ on compact $\mathbb S$ give
$0<K_{\min}\leq K(s)\leq K_{\max}<\infty$, while continuity of $f$
on $[0,b]\times\mathbb S$ gives local boundedness for each
$b<\infty$. Corollary~\ref{cor:exact_tail_mapping} therefore
applies with $\rho=1$ and $H=K$.

\newpage

\section{Model coverage of the impossibility result}
\label{app:survey}

This appendix surveys heterogeneous-agent models associated with
heavy wealth tails and asks whether their optimal consumption policies
satisfy the positive linear lower bound in
Assumption~\ref{a:1}. The mechanisms generating these wealth tails
are summarized in Table~\ref{tab:wealth_mechanisms}. In the
permanent-income row, $G'$ denotes next-period gross
permanent-income growth and the multiplier applies to wealth
normalized by permanent income, and in the persistent-discount
row, $\beta$ denotes the discount factor. The multipliers are
high-wealth coefficients in the sense of
Section~\ref{sec:ha_models}: whether a stationary Pareto tail
exists depends on the relevant process and hypotheses there, with
the exact-affine matrix of equation~\eqref{eq:kesten_matrix} and
the expansion-only matrix of equation~\eqref{eq:mst_matrix} kept
distinct.

\begin{table}[htbp]
\centering
\footnotesize\renewcommand{\arraystretch}{0.95}\setlength{\abovecaptionskip}{4pt}
\caption{Main mechanisms for Pareto-tailed stationary wealth}
\label{tab:wealth_mechanisms}
\begin{tabular}{@{}p{2.9cm}p{4.6cm}p{3.0cm}p{2.7cm}@{}}
\toprule
Mechanism & Economic interpretation & Multiplier & Reference \tabularnewline
\midrule
\raggedright Stochastic idiosyncratic returns\par &
\raggedright Portfolio risk, entrepreneurial income, private business
returns\par & $A=R'\,s_\infty(S)$ &
\raggedright \citet{benhabib2015wealth}\par \tabularnewline
\raggedright Persistent discount heterogeneity\par &
\raggedright Patient households save more persistently than impatient
households\par & $A=R\,s_\infty(S,\beta)$ &
\raggedright \citet{toda2019wealth}\par \tabularnewline
\raggedright Permanent-income normalization\par &
\raggedright Dividing by idiosyncratic permanent-income growth makes the
effective return stochastic\par & $A=(R/G')\,s_\infty(S)$ &
\raggedright \citet{carroll1997buffer,ma2020income}\par \tabularnewline
\raggedright Dynastic or bequest scaling\par &
\raggedright Intergenerational transmission creates multiplicative wealth
propagation\par & model specific &
\raggedright \citet{benhabib2011distribution}\par \tabularnewline
\bottomrule
\end{tabular}
\end{table}

Tables~\ref{tab:survey_a1}, \ref{tab:survey_a2}, and
\ref{tab:survey_b} summarize the findings. In the tables, the column ``Trapped?'' records whether the
paper's lower-bound logic restricts the consumption tail. When
wealth has a Pareto-type index, Theorem~\ref{thm:3} rules out
$\alpha_C > \alpha_W$. For the nonlinear income-fluctuation problem,
where \citet{ma2020income} establish a Pareto-order lower bound
rather than an exact wealth-tail index,
Theorem~\ref{thm:trap} instead rules out an exact
consumption-tail index above the Pareto-order boundary $\kappa_W$. A
$\checkmark$ means Assumption~\ref{a:1} is established by a
formal result in the cited paper or follows by direct specialization
from such a result. A $\checkmark^*$ means
Assumption~\ref{a:1} is supported by an economic argument (the
high-wealth homogeneity reasoning stated below) but has not, to our
knowledge, been formally proved in the cited paper. A $\times$ means
Assumption~\ref{a:1} fails by a formal result in the cited
paper. A reader can therefore distinguish entries where the
impossibility or its failure is a theorem from entries where the
status is a well-motivated conjecture or open.

Under an exact Pareto wealth tail, independence of stationary
wealth $W$ from the exogenous state $S$, and the affine or concave
asymptotically linear policy conditions of
Proposition~\ref{th:2}, that proposition sharpens the
impossibility to the exact equality
$\alpha_C = \alpha_W$. Appendix~\ref{app:discussion} records when
the independence condition holds and the weaker tail-factorization
alternative.

\paragraph{The high-wealth homogeneity argument.}
Several table entries marked $\checkmark^*$ rely on the following
reasoning, stated once to avoid repetition. Under CRRA utility, the
budget constraint~\eqref{eq:budget} becomes approximately
homogeneous of degree one in wealth $w$ as $w \to \infty$, because
nonfinancial income $Y$ has finite mean
(Assumption~\ref{ass:existence}(iv)) and is therefore negligible
relative to $w$. The value function inherits this homogeneity,
behaving like $w^{1-\gamma}/(1-\gamma)$ times a state-dependent
factor at large $w$, and optimal consumption is approximately
$f(w,s) \approx m_\infty(s)\,w$ with $m_\infty(s) > 0$. This
argument is formalized for the discrete-time income-fluctuation
problem by \citet[Proposition~2.5]{ma2020income}. Under the
conditions of Corollary~\ref{cor:trap_hara}, shifted HARA
utility retains the linear lower bound, although existence of the
asymptotic consumption slope is not established there. For models not
directly covered by these results, a formal proof would require
verifying the relevant concavity-preservation condition
\citep[condition~(17)]{ma2020income} or its continuous-time
analogue. The $\checkmark^*$ entries below record that this step
remains conjectural in the cited settings.

\begin{table}[htbp]
\centering
\footnotesize\renewcommand{\arraystretch}{0.92}\setlength{\abovecaptionskip}{4pt}
\caption{Core frameworks: does the impossibility
apply?}\label{tab:survey_a1}
\begin{tabular}{@{}p{4.8cm}p{2.3cm}p{1.6cm}p{5.5cm}@{}}
\toprule
Model class & \centering Heavy wealth tail?\par & \centering
Trapped?\par & \raggedright Why Assumption~\ref{a:1}
holds\par \tabularnewline
\midrule
\raggedright Risk-free saving, constant $\beta$, under the conditions of
Stachurski and Toda
\citep{bewley1976permanent,huggett1993risk,
aiyagari1994uninsured}\par
  & \centering inherits the income tail \newline (Thms.~3, 8)\par
  & \centering $\checkmark$\par
  & \raggedright Prop.~5 of \citet{stachurski2019impossibility}, as
  corrected in their 2020 corrigendum, gives the positive linear
  consumption lower bound\par \tabularnewline
\raggedright General-equilibrium Bewley economy with i.i.d.\
stochastic returns, CRRA
\newline \citep{benhabib2015wealth}\par
  & \centering yes \newline (their Thm.~4)\par
  & \centering $\checkmark$\par
  & \raggedright $f$ asymptotically linear with
  $m_\infty \in (0,1)$
  \citep[Prop.~5]{benhabib2015wealth}\par \tabularnewline
\raggedright Stochastic returns, Markov
\newline \citep{ma2020income}\par
  & \centering lower bound \newline (their Thm.~3.3)\par
  & \centering $\checkmark$\par
  & \raggedright $f(w,s) \geq (1-\bar{x})w$
  \citep[Prop.~2.6]{ma2020income} and is
  asymptotically linear
  under CRRA
  \citep[Prop.~2.5]{ma2020income}\par \tabularnewline
\raggedright Overlapping generations + stochastic returns
\newline
\citep{benhabib2011distribution}\par
  & \centering yes \newline (their Thms.~1--2)\par
  & \centering $\checkmark$\par
  & \raggedright $f$ affine in wealth:
  $f(w,s) = m(s)(w + p(s))$
  \citep[Appendix~A]{benhabib2011distribution}.
  Proposition~\ref{prop:olg} transfers the population power
  bound to consumption\par \tabularnewline
\raggedright Random discount factors
\newline \citep{toda2019wealth}\par
  & \centering yes \newline (their Sec.~3)\par
  & \centering $\checkmark$\par
  & \raggedright $f(w,s)=a_s^{-1/\gamma}w$
  \citep[Prop.~2.1]{toda2019wealth}\par \tabularnewline
\bottomrule
\end{tabular}
\end{table}

\begin{table}[htbp]
\centering
\footnotesize\renewcommand{\arraystretch}{0.92}\setlength{\abovecaptionskip}{4pt}
\caption{Extensions: does the impossibility apply?}
\label{tab:survey_a2}
\begin{tabular}{@{}p{4.8cm}p{2.3cm}p{1.6cm}p{5.5cm}@{}}
\toprule
Model class & \centering Heavy wealth tail?\par & \centering
Trapped?\par & \raggedright Why Assumption~\ref{a:1}
holds\par \tabularnewline
\midrule
\raggedright Shifted HARA utility \newline
\citep{ma2022asymptotic}\par
  & \centering conditional on a separate wealth-tail assumption\par
  & \centering $\checkmark$\par
  & \raggedright Corollary~\ref{cor:trap_hara} delivers the
  linear lower bound under the \citet{ma2022asymptotic} foundation
  and the stability condition, without policy concavity\par \tabularnewline
\raggedright Perpetual youth + returns
\newline \citep{benhabib2016distribution}\par
  & \centering yes \newline (their Prop.~2)\par
  & \centering $\checkmark$\par
  & \raggedright $C(s,t)=A^{-1/\gamma}W(s,t)$
  \citep[Prop.~1]{benhabib2016distribution}\par \tabularnewline
\raggedright Persistent heterogeneous returns, continuous time
\newline \citep{cao2017persistent}\par
  & \centering yes\par
  & \centering $\checkmark^*$\par
  & \raggedright CRRA, with the high-wealth homogeneity
  argument\par \tabularnewline
\raggedright Stochastic returns + random
$\beta$ \newline
\citep{hubmer2021sources}\par
  & \centering yes\par
  & \centering $\checkmark^*$\par
  & \raggedright Saving rules documented as
  asymptotically linear
  (numerical, not proved
  formally)\par \tabularnewline
\raggedright Scale-dependent returns
extension
\newline \citep{gaillard2023consumption}\par
  & \centering yes\par
  & \centering $\checkmark^*$\par
  & \raggedright Their Prop.~1, for homothetic preferences without
  a wealth motive and under its parameter restriction, gives
  $c\sim b_n w^{1+\eta}$ with $b_n>0$: for $\eta>0$ the policy is
  superlinear at high wealth, so the linear lower bound holds
  state by state, with a single threshold uniform across
  labor-productivity states the remaining step. Their mapping
  $\alpha_C=\alpha_W/(1+\eta)\leq\alpha_W$ is consistent with
  Theorem~\ref{thm:3}, and the scale-independent case ($\eta=0$)
  gives equality (Appendix~\ref{app:discussion})\par \tabularnewline
\bottomrule
\end{tabular}
\end{table}

\begin{table}[htbp]
\centering
\footnotesize\renewcommand{\arraystretch}{0.92}\setlength{\abovecaptionskip}{4pt}
\caption{Models that may escape the impossibility.}
\label{tab:survey_b}
\begin{tabular}{@{}p{4.8cm}p{2.3cm}p{1.6cm}p{5.5cm}@{}}
\toprule
Model class & \centering Heavy wealth tail?\par & \centering
Trapped?\par & \raggedright Why Assumption~\ref{a:1} may
fail\par \tabularnewline
\midrule
\raggedright Preference for wealth
($\delta < \gamma$) \newline
\citep{ma2025optimal}\par
  & \centering in their calibrated stochastic-return model\par
  & \centering $\times$\par
  & \raggedright $m_\infty = 0$, so
  Assumption~\ref{a:1} fails. Their Theorem~2 gives the policy
  exponent $\delta/\gamma$ \citep[Thm.~2]{ma2025optimal}. Given a
  weak Pareto-type wealth index and the conditions of
  Theorem~\ref{thm:tail_mapping}, the weak consumption index is
  $\alpha_W\gamma/\delta$. See
  \citet[Chapter~5]{hu2026pareto}\par \tabularnewline
\raggedright Zero asymptotic MPC under
high $\gamma$ \newline
\citep{ma2021theory}\par
  & \centering their calibrated return-risk example\par
  & \centering \textbf{possibly not}\par
  & \raggedright In their calibrated return-risk example, sufficiently
  high risk aversion yields a zero asymptotic MPC. This result alone
  leaves the consumption tail index open.\par \tabularnewline
\raggedright Multi-asset / portfolio
\newline \citep{kaplan2014model,
gomez2025wealth}\par
  & \centering open\par
  & \centering \textbf{open}\par
  & \raggedright \citet[Chapter~5]{hu2026pareto} proves an escape for
  trade-size liquidation costs whose induced effective
  consumption cost is regularly varying with index above
  one under its
  high-wealth reduction condition. Holdings-scaled convexity alone
  does not deliver that result, and the generic multi-asset case
  remains unresolved\par \tabularnewline
\bottomrule
\end{tabular}
\end{table}

\newpage

\section{Related results, benchmarks, and open problems}
\label{app:discussion}

\subsection*{Relationship to Stachurski--Toda and Gaillard et al.}

Two prior results are most closely related to the impossibility
theorem.

\paragraph{Stachurski--Toda.}
{\sloppy
\citet{stachurski2019impossibility} prove that in
Bewley--Huggett--Aiyagari models with
infinitely-lived agents, risk-free saving ($R$ constant), and
constant discount factors ($\beta$ constant), the wealth
distribution inherits the tail behavior of income shocks (their
Theorems~3 and~8). Their impossibility concerns wealth versus
income, whereas ours concerns wealth versus consumption. A direct
connection exists. Under its strengthened assumptions,
Proposition~5$'$ of \citet{stachurski2020corrigendum} establishes
that optimal consumption satisfies $c(a,z)\geq ma$ for a constant
$m>0$. In our notation, this is $f(w,s)\geq mw$, exactly
Assumption~\ref{a:1} of Theorem~\ref{thm:3} with lower-bound
constant $m$ and $w_0=0$. Thus, under the corrigendum's
assumptions, the \citet{stachurski2019impossibility} setting
satisfies the policy condition used here. The related result
of \citet{benhabib2017earnings}, which shows that a class of OLG
models cannot jointly match earnings and wealth tails, also
operates through the linear structure of optimal consumption
\citep[Appendix~A]{benhabib2011distribution}.\par}

\paragraph{Gaillard et al.}
The continuous-time model of \citet{gaillard2023consumption} combines
CRRA utility, stochastic capital income, and perpetual-youth
demographics. In their model, consumption and wealth share the same Pareto
tail index under homothetic preferences and scale-independent
returns \citep[Proposition~1]{gaillard2023consumption}.
Their derivation relies on two features of their scale-independent
benchmark ($\eta=0$), for homothetic preferences without a wealth
motive and under the parameter restriction of their Proposition~1:
(i) the consumption policy is asymptotically linear in wealth (a
consequence of CRRA utility and the homogeneous HJB equation in
their perpetual-youth setting), and (ii) a tail-factorization property of the joint
distribution of wealth and the exogenous state. Property~(i)
delivers Assumption~\ref{a:1} along each state, with uniformity
across states the remaining step, exactly as the qualified survey
row records (Appendix~\ref{app:survey}). Theorem~\ref{thm:3}
requires neither feature: only Assumption~\ref{a:1} and a
Pareto-type tail for wealth. The open problems below
state the factorization condition and the open problem of
extending Proposition~\ref{th:2}. Appendix~\ref{app:survey}
records, model by model, whether the positive linear lower bound is
established formally or supported by the high-wealth homogeneity
argument across the heterogeneous-agent literature.

\subsection*{Exact affine and multiplicative benchmarks}

For an exact affine recursion
$W_{t+1}=A_{t+1}W_t+B_{t+1}$ with Markov-dependent coefficients,
Markov extensions of Kesten--Goldie theory establish power-law tails
under additional moment, irreducibility, and nonarithmetic conditions
\citep{roitershtein2007one}. In the finite-state benchmark used here,
the multiplier moments are collected in
$K_\alpha$~\eqref{eq:kesten_matrix}, with entries
$Q_{ij}\EE[(s_\infty(s_i)R(s_j,\zeta))^\alpha]$. Its spectral crossing
is the tail index used in the exact-affine calibration.

\citet{beare2022determination} study a different exact process:
Markov multiplicative growth with state-dependent random resets. Their
Pareto tail index also solves a spectral-radius equation, but that
equation includes the reset or survival probabilities as well as the
multiplier moments. The nonlinear income-fluctuation recursion is
different again. Theorem~3.3 of \citet{ma2020income} uses the
expansion-only matrix $K^+_\kappa$ in~\eqref{eq:mst_matrix} to
establish the Pareto-order lower bound in~\eqref{eq:mst_tail}.
These spectral conditions are not interchangeable: $K_\alpha$ gives
the exact-affine tail index, $K^+_\kappa$ only the Pareto-order lower
bound for the nonlinear model, and the two need not coincide. An exact-tail result for the nonlinear
recursion would require additional arguments.

\subsection*{Open problems}

\paragraph{General utility.}
Proposition~\ref{prop:stab} delivers the linear lower bound
$f(w,s) \geq (1-\bar{x})w$ for any $u$ satisfying
Assumption~\ref{ass:existence}, without requiring CRRA or HARA.
Corollary~\ref{cor:trap_hara} delivers the same bound for
shifted HARA utility, whose finite marginal utility at zero places
it outside Assumption~\ref{ass:existence}, and its argument
uses no policy concavity. The Pareto trap
(Theorem~\ref{thm:trap}) additionally requires the limit
$m_\infty(s) \in (0,1)$, which is established under CRRA
(Corollary~\ref{cor:trap_crra}).
Remark~\ref{rem:hara_scope} states what the shifted-HARA
extension does and does not establish. Verifying the
concavity-preservation condition
\citep[condition~(17)]{ma2020income} for the shifted-HARA class is
the natural attack.

\paragraph{Tail factorization.}
Proposition~\ref{th:2} requires independence of stationary wealth
$W$ and the exogenous state $S$. In the i.i.d.\ benchmark of
\citet{benhabib2015wealth}, there is no nondegenerate exogenous policy
state, so the condition is automatic. In OLG models where each
generation's initial state is drawn independently of inherited
wealth \citep{benhabib2011distribution}, it holds at birth, while
at later ages current wealth and the within-life state are
generally dependent in the pooled population, so the condition
must be verified for the population actually compared.
I.i.d.\ innovations alone do not
guarantee this independence under every timing convention, and it need
not hold in Markov settings \citep{ma2020income}.
\citet{gaillard2023consumption} use a weaker
condition: multiplicative separability of the joint density of
$(W,S)$ in the tail (their footnote~27), under which the conditional
distribution of $S$ given $\{W>w\}$ converges as $w\to\infty$.
An extension of Proposition~\ref{th:2} to such tail factorization
could cover their asymptotically linear consumption case if the other
policy and tail assumptions also hold.

\paragraph{Joint tails.}
Theorem~\ref{thm:3} and Proposition~\ref{th:2} concern marginal
tail indices. The dependence structure between wealth and
consumption in the tail is a separate object, characterized
by quantities such as the upper-tail dependence coefficient and the
stable tail dependence function. Whether the impossibility logic
extends to these joint-tail objects is open.

\paragraph{Policy wedges and aggregate risk.}
{\sloppy
Whether progressive taxation or aggregate shocks can violate the
positive linear lower bound remains open. The benchmark includes
neither feature.
Subsection~\ref{ssec:escape_criterion} states that condition
precisely.\par}

\paragraph{Endogenous labor supply and portfolio choice.}
Endogenous labor supply and multi-asset portfolio choice remain open
beyond the mechanisms studied in \citet[Chapter~5]{hu2026pareto}. In
multi-asset models \citep{kaplan2014model,gomez2025wealth},
consumption depends on the composition of wealth, not just its
level. Under the high-wealth reduction in \citet[Chapter~5]{hu2026pareto},
a trade-size cost with effective-cost index above one yields sublinear
consumption. In the one-asset benchmark, an asymptotically linear
effective cost preserves the weak tail index under the chapter's
stationarity and positive limiting expenditure-share assumptions.
For the holdings-scaled cost considered there, cost is proportional
to the position at a fixed trade share. This scaling alone leaves
the optimal consumption policy open. Whether other portfolio
structures generate sublinearity remains open.
Similarly, if labor supply responds to wealth, the effective
return on saving may change at the top, potentially altering the
asymptotic form of the consumption rule.

\paragraph{Continuous-time formal proofs.}
The high-wealth homogeneity argument suggests that consumption is
asymptotically linear in continuous-time HA models with stochastic
returns \citep{gabaix2016power,cao2017persistent}. Establishing this
claim under explicit conditions, through a proof analogous to
\citet[Proposition~2.5]{ma2020income}, would place these models on the
same formal footing as the discrete-time cases.

\section{Remarks, data, methodology and robustness checks}
\label{app:empirics}

\subsection{Remarks on Pareto tail definition}
\label{app:remarks}

\begin{figure}[ht]
    \centering
    \begin{minipage}[t]{0.45\textwidth}
        \centering
        \includegraphics[width=\textwidth]{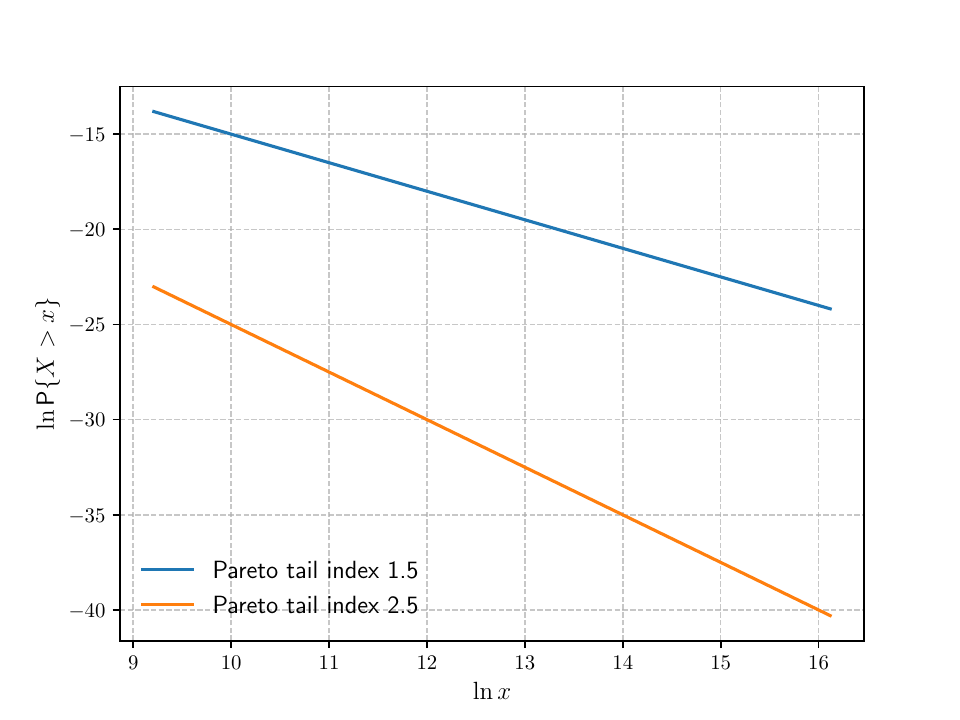}
        \caption{Log-log CCDFs: Pareto tails with indices $1.5$ and $2.5$}
        \label{fig:plot1}
    \end{minipage}
    \hfill
    \begin{minipage}[t]{0.45\textwidth}
        \centering
        \includegraphics[width=\textwidth]{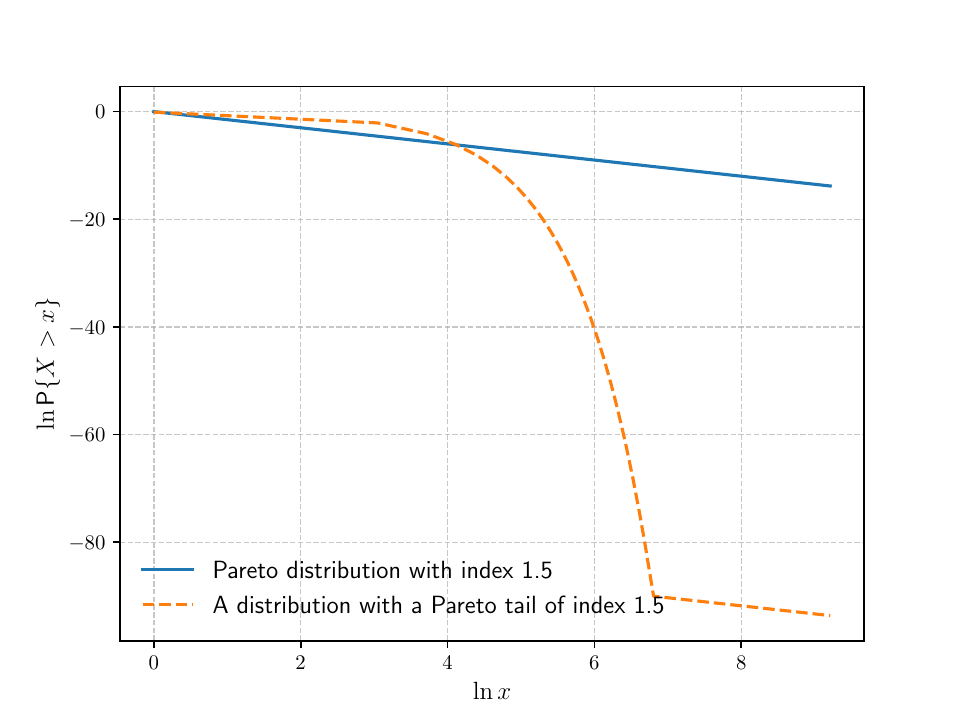}
        \caption{A distribution with a Pareto tail (index $1.5$) that is not globally Pareto}
        \label{fig:plot2}
    \end{minipage}
\end{figure}

\begin{remark}
    In practice, for large values of $x$, we plot $\ln \bar F_X(x)$ against $\ln x$ (a \emph{log-log} or \emph{double-log plot}) to assess visually whether $X$ exhibits a Pareto tail. By Definition~\ref{df:01}, if $F_X$ has a Pareto tail with index $\alpha > 0$, then for some constant $k > 0$ and sufficiently large $x$, $\bar F_X(x) \simeq k x^{-\alpha}$. Taking logarithms gives $\ln \bar F_X(x) \simeq \ln k - \alpha \ln x$, so the CCDF appears as a straight line in the double-log plot. Figure~\ref{fig:plot1} illustrates this for two Pareto tails with indices $1.5$ and $2.5$.
\end{remark}

\begin{remark}
    The Pareto tail index measures the heaviness of a distribution's tail. As illustrated in Figure~\ref{fig:plot1}, a lower index indicates a heavier tail: extremely large values are more prevalent, a greater share of the distribution is concentrated among the top percentile, and inequality at the top is higher.
\end{remark}

\begin{remark}
    A Pareto tail does not imply a Pareto distribution. Figure~\ref{fig:plot2} shows a distribution $F_X$ that follows an exponential distribution with rate $\lambda = 0.1$ for $0 < x < x_m = 900$ and a Pareto tail with index $\alpha = 1.5$ for $x \geq x_m$: specifically, $\bar F_X (x) = k x^{-\alpha}$ with $k := \exp(-\lambda x_m) x_m^{\alpha}$ for $x \geq x_m$, alongside a pure Pareto distribution with the same index.
\end{remark}

\subsection{Eligibility and multiple comparisons}
\label{app:eligibility}

An index pair enters asymptotic comparisons only if it passes four checks fixed before estimation.
\begin{enumerate}
    \item \emph{Construction.} The consumption and wealth variables pass the audited construction rules on the matched household sample, with documented treatment of missing values.
    \item \emph{Fit.} The Pareto model is an adequate description above the fixed threshold. Adequacy is assessed by a bootstrap goodness-of-fit test at the released threshold, and a failing margin disqualifies the pair.
    \item \emph{Threshold stability.} The fitted index is stable as the threshold moves across the prespecified quantile window from the 85th to the 95th percentile.
    \item \emph{Inference.} The design-based uncertainty is well defined, with at least 100 raw and 50 effective tail observations behind each fit.
\end{enumerate}
Ineligible pairs remain in the family with a p-value of one. The seven one-sided tests are combined by Holm's step-down rule \citep{holm1979simple}. Order the raw p-values from smallest to largest, compare the smallest to $0.05/7$, the next to $0.05/6$, and so on, and stop at the first failure. The rule holds the familywise error rate at 5 percent whatever the dependence between the tests. The exact-tail benchmark is stricter than eligibility. It additionally requires an eligible Pareto fit on both margins with both indices above one, and it is the only status that licenses calculations assuming an exact Pareto distribution.

\subsection{Data}
\label{app:data}

The data used in this study include waves 2, 6--10, 14, and 18 of the Household, Income and Labour Dynamics in Australia (HILDA) survey and the 2007, 2013, and 2018 waves of the Chinese Household Income Project Survey (CHIPS). HILDA is an annual survey conducted in Australia. CHIPS collects similar information in China. These datasets were selected because they are among the few that provide both wealth and consumption information in their respective countries.

We define wealth and consumption at the household level as follows. Wealth is net worth: the difference between total assets and total liabilities. Consumption in HILDA is the sum of all nondurable goods expenditures available in each wave. For CHIPS, consumption is defined using the consumption expenses variable (\textit{n4202}), as specified in the survey.

Using these definitions, we construct household-level distributions of wealth and consumption for each wave in HILDA and CHIPS through aggregation and deflation. We also construct individual-level distributions for HILDA by applying adult equivalization. Details of the aggregation, deflation, and equivalization processes are in Appendix~\ref{app:ad}.

\begin{table}[H]
    \centering
    \resizebox{0.7\linewidth}{!}{
    \begin{tabular}{lrrrrrr}
        \toprule
        Wave number & 6 & 7 & 8 & 9 & 10 & pooled \\
        \midrule
        Sample size & 7139 & 7063 & 7066 & 7234 & 7317 & 35819 \\
        Min         & 0.00 & 7.67 & 0.00 & 0.00 & 0.00 & 0.00 \\
        Max         & 2432.5 & 1950.8 & 5080.4 & 2928.1 & 2440.8 & 5080.4 \\
        Mean        & 143.8 & 147.5 & 151.3 & 144.0 & 148.1 & 147.0 \\
        Median      & 122.2 & 125.1 & 127.5 & 124.0 & 125.8 & 124.9 \\
        Standard deviation & 108.3 & 104.2 & 125.0 & 98.2 & 110.9 & 109.7 \\
        Skewness    & 6.04 & 4.75 & 12.24 & 6.24 & 6.46 & 7.82 \\
        Excess kurtosis & 74.16 & 45.72 & 376.58 & 112.85 & 88.04 & 180.57 \\
        p90         & 236.6 & 246.4 & 250.2 & 239.7 & 245.5 & 243.7 \\
        p95         & 296.1 & 303.6 & 319.3 & 302.0 & 304.4 & 305.0 \\
        p99         & 522.2 & 515.9 & 558.3 & 466.1 & 531.1 & 518.5 \\
        Top 10\% share   & 0.252 & 0.247 & 0.256 & 0.241 & 0.250 & 0.249 \\
        Top 1\% share    & 0.0584 & 0.0546 & 0.0612 & 0.0491 & 0.0572 & 0.0561 \\
        Top 0.1\% share  & 0.0130 & 0.0109 & 0.0147 & 0.0109 & 0.0136 & 0.0119 \\
        \bottomrule
    \end{tabular}
    }
    \caption{Descriptive statistics of HILDA nondurable consumption (deflated and adult-equivalized). Kurtosis reported is excess kurtosis (i.e., minus 3).}
    \label{table:tb5}
\end{table}

\begin{table}[H]
    \centering
    \resizebox{0.7\linewidth}{!}{
    \begin{tabular}{lrrrrrr}
        \toprule
        Wave number & 2 & 6 & 10 & 14 & 18 & pooled \\
        \midrule
        Sample size & 6889 & 6784 & 6940 & 8978 & 9116 & 38707 \\
        Min & 0.0039 & 0.0053 & 0.0074 & 0.0038 & 0.0039 & 0.0038 \\
        Max & 47516 & 61566 & 60008 & 57259 & 59411 & 61566 \\
        Mean & 2364.0 & 3223.3 & 3172.5 & 3189.6 & 3619.6 & 3146.8 \\
        Median & 1288.4 & 1710.9 & 1713.0 & 1724.3 & 1970.3 & 1673.2 \\
        Standard deviation & 3767.3 & 5702.9 & 5232.1 & 5094.4 & 5475.8 & 5135.8 \\
        Skewness & 4.93 & 5.31 & 5.09 & 4.80 & 4.17 & 4.91 \\
        Excess kurtosis & 35.90 & 37.73 & 37.43 & 34.22 & 25.81 & 35.15 \\
        p90 & 5283.3 & 7018.3 & 7272.5 & 7491.2 & 8618.5 & 7218.8 \\
        p95 & 8005.9 & 10744.8 & 10656.8 & 11065.2 & 12523.8 & 10872.8 \\
        p99 & 19297.5 & 39234.6 & 29382.0 & 25335.2 & 29089.8 & 27789.3 \\
        Top $10\%$ share & 0.451 & 0.480 & 0.459 & 0.454 & 0.444 & 0.46 \\
        Top $1\%$ share  & 0.121 & 0.138 & 0.127 & 0.121 & 0.110 & 0.12 \\
        Top $0.1\%$ share & 0.01937 & 0.01971 & 0.01908 & 0.01800 & 0.01801 & 0.019 \\
        \bottomrule
    \end{tabular}}
    \caption{Descriptive statistics of HILDA net worth (deflated and adult-equivalized). Kurtosis reported is excess kurtosis (i.e., minus 3).}
    \label{table:tb6}
\end{table}

The descriptive evidence already points toward thinner consumption tails and heavier wealth tails.

\begin{table}[H]
    \centering
    \resizebox{0.6\linewidth}{!}{
    \begin{tabular}{lrrrr}
        \toprule
        Wave number & 2007 & 2013 & 2018 & pooled \\
        \midrule
        Sample size & 15746 & 18948 & 10754 & 45448 \\
        Min & 0.0010 & 0.0064 & 0.01 & 0.0010 \\
        Max & 7560 & 6900 & 3240 & 7560 \\
        Mean & 90.54 & 156.49 & 195.22 & 139.67 \\
        Median & 67.38 & 119.35 & 133.49 & 95.41 \\
        Standard deviation & 96.81 & 141.32 & 168.72 & 138.56 \\
        Skewness & 14.24 & 7.08 & 3.45 & 8.73 \\
        Excess kurtosis & 577.62 & 126.45 & 22.26 & 178.60 \\
        p90 & 179.88 & 302.61 & 371.43 & 270.03 \\
        p95 & 229.30 & 394.10 & 506.53 & 367.62 \\
        p99 & 406.97 & 754.84 & 908.37 & 706.47 \\
        Top 10\% share & 0.30 & 0.29 & 0.29 & 0.30 \\
        Top 1\% share & 0.068 & 0.069 & 0.070 & 0.073 \\
        Top 0.1\% share & 0.014 & 0.014 & 0.013 & 0.014 \\
        \bottomrule
    \end{tabular}
    }
    \caption{Descriptive statistics of CHIPS consumption (deflated)}
    \label{table:tb11}
\end{table}

\begin{table}[H]
    \centering
    \resizebox{0.6\linewidth}{!}{
    \begin{tabular}{lrrr}
        \toprule
            Wave number & & Pooled (2013 \& 2018) & \\
            Wealth & Rural & Urban & Total \\
        \midrule
        Sample size & 10406 & 13309 & 23715 \\
        Min & 0.74 & 0.37 & 0.37 \\
        Max & 342468 & 217068 & 342468 \\
        Mean & 928 & 1736 & 1381 \\
        Median & 431.64 & 813.16 & 599.45 \\
        Standard deviation & 3989 & 4564 & 4340 \\
        Skewness & 63.26 & 20.52 & 34.98 \\
        Excess kurtosis & 5211.57 & 690.8 & 2101.4 \\
        p90        & 1921.2 & 3648.8  & 2904.2      \\
        p95        & 2893.1 & 5555.6  & 4513.2    \\
        p99        & 6886.68 & 15266.16  & 11974.01  \\
        Top $10\%$ share        & 0.503 & 0.50  & 0.52      \\
        Top $1\%$ share        & 0.198 & 0.182  & 0.196     \\
        Top $0.1\%$ share        & 0.087 & 0.061  & 0.069      \\
        \bottomrule
    \end{tabular}
    }
    \caption{Descriptive statistics of CHIPS pooled wealth ($W>0$)}
    \label{table:tb12}
\end{table}

The Chinese results reinforce the Australian pattern.

\subsection{Processing the data: aggregation, deflation and equivalization}
\label{app:ad}

\subsubsection{Aggregation}
\label{app:agg}

We construct net worth and consumption for each household by aggregating the corresponding micro-data.

In the HILDA survey, only waves 2, 6, 10, 14, and 18 (denoted by the letters \textit{b}, \textit{f}, \textit{j}, \textit{n}, and \textit{r} respectively) contain wealth-related variables. The variables \textit{$\chi$hwnwip} and \textit{$\chi$hwnwin} represent the positive and negative parts of household net worth, respectively, where $\chi$ is replaced with the letter corresponding to the wave. A household's signed net worth is their difference, \textit{$\chi$hwnwip} minus \textit{$\chi$hwnwin}.

Only HILDA waves from 2006 to 2010 include comprehensive lists of both nondurable and durable consumption variables. Table~\ref{table:tb14} summarizes all variables representing nondurable and durable consumption. For each household, nondurable consumption is the sum of all nondurable variables, and total consumption is the sum of all nondurable and durable variables listed in Table~\ref{table:tb14}.

\begin{table}[H]
    \centering
    \resizebox{0.45\linewidth}{!}{
    \begin{tabular}{lll}
        \hline
        Nondurables & Nondurables & Durables \\
        \hline
        \textit{$\chi$hxygrci} & \textit{$\chi$hxytlii} & \textit{$\chi$hxyncri} \\
        \textit{$\chi$hxyalci} & \textit{$\chi$hxyholi} & \textit{$\chi$hxyucri} \\
        \textit{$\chi$hxycigi} & \textit{$\chi$hxyphii} & \textit{$\chi$hxycmpi} \\
        \textit{$\chi$hxypbti} & \textit{$\chi$hxyoii} & \textit{$\chi$hxytvi} \\
        \textit{$\chi$hxymli} & \textit{$\chi$hxyhlpi} & \textit{$\chi$hxywgi} \\
        \textit{$\chi$hxymvfi} & \textit{$\chi$hxyphmi} & \textit{$\chi$hxyfrni} \\
        \textit{$\chi$hxymcfi} & \textit{$\chi$hxyutli} & \\
        \textit{$\chi$hxywcfi} & \textit{$\chi$hxyhmri} & \\
        \textit{$\chi$hxyccfi} & \textit{$\chi$hxymvri} & \\
        \textit{$\chi$hxyedci} & & \\
        \hline
    \end{tabular}
    }
    \tablefootnote{$\chi$ should be replaced by the letter representing the corresponding waves 6--10.}
    \caption{Consumption variables in HILDA}
    \label{table:tb14}
\end{table}

In the CHIPS survey, only waves from 2007, 2013, and 2018 provide comprehensive consumption information. For each household, consumption is defined using the consumption expenses variable (\textit{n4202}), and wealth is the difference between total assets and total liabilities. Table~\ref{table:tb15} lists all asset and debt variables for both urban and rural households in CHIPS.

For an urban household, net worth is the difference between total assets (the sum of all variables in column~1 of Table~\ref{table:tb15}) and total liabilities (column~3). For a rural household, net worth is defined analogously using column~2 for assets and column~3 for liabilities. Pooling urban and rural households for each wave yields the household-level wealth distribution.

\begin{table}[H]
    \centering
    \resizebox{0.5\linewidth}{!}{
    \begin{tabular}{lll}
        \hline
        Urban assets & Rural assets & Urban/Rural debt \\
        \hline
        \textit{f03} & \textit{f03} & \textit{f05} \\
        \textit{f04} & \textit{f04} & \\
        \textit{f08} & \textit{f08} & \\
        \textit{f07\_1} & \textit{f07\_1} & \\
        \textit{f07\_2} & \textit{f06\_1} & \\
        \textit{f06\_1} & \textit{f06\_2} & \\
        \textit{f06\_2} & \textit{f06\_3} & \\
        \textit{f06\_3} & \textit{f06\_4} & \\
        \textit{f06\_4} & \textit{f06\_5} & \\
        \textit{f06\_5} & \textit{f06\_6} & \\
        \textit{f06\_6} & & \\
        \hline
    \end{tabular}
    }
    \caption{Assets and debt variables for urban and rural households in CHIPS}
    \label{table:tb15}
\end{table}

\subsubsection{Deflation}
\label{app:def}

We deflate the net worth and consumption values for each household in each wave of HILDA and CHIPS to make values across waves comparable, enabling construction of pooled distributions. We use the OECD ``Consumer Price Index: Total'' as the price index for both HILDA (Australia, variable \textit{AUSCPIALLAINMEI} in FRED\footnote{See \href{https://fred.stlouisfed.org/series/AUSCPIALLAINMEI}{https://fred.stlouisfed.org/series/AUSCPIALLAINMEI}}) and CHIPS (China, variable \textit{CHNCPIALLMINMEI} in FRED\footnote{See \href{https://fred.stlouisfed.org/series/CHNCPIALLMINMEI}{https://fred.stlouisfed.org/series/CHNCPIALLMINMEI}}). We normalize the price level to 100 in the year 2000 for Australia and 2007 for China, then divide each wave's values by the corresponding normalized price. This yields deflated household-level distributions of net worth and consumption for each wave.

\subsubsection{Equivalization}
\label{app:equiv}

To obtain individual-level distributions, we divide household wealth or consumption by the number of adult equivalents in the household. Following \cite{kocherlakota2009asset}, the number of adult equivalents is $(A + \theta_1 K)^{\theta_2}$, where $A$ is the number of adults, $K$ is the number of children, and $\theta_1, \theta_2$ are parameters. Consistent with the existing literature, we set $\theta_1 = 0.7$ and $\theta_2 = 0.65$.

We apply this step only to HILDA data, where the household adult count (variable \textit{$\chi$hhadult}) and total household size (\textit{$\chi$hhpers}) are directly available. The number of children is imputed as \textit{$\chi$hhpers} $-$ \textit{$\chi$hhadult}. Individual net worth or consumption is then household value divided by the number of adult equivalents.

\subsection{Estimation approach and hypothesis testing}
\label{app:ea}

This section provides a brief overview of the Maximum Likelihood Estimation (MLE) method and the hypothesis testing approach used in \cite{clauset2009power}.

Let $X$ be a random variable with distribution $F_X$ and complementary CDF $\bar F_X(x) := 1 - F_X(x)$. Assume that $F_X$ has a Pareto tail with index $\alpha > 0$ on $[x_{\min}, \infty)$ for some sufficiently large $x_{\min}$ (the \emph{lower bound} of the Pareto tail). Specifically, for $x \geq x_{\min}$, $\bar F_X (x) \simeq k x^{-\alpha}$ for some constant $k > 0$.

Let $x_i$, $i = 1, \dots, n$, be independently and identically distributed observations from $X$ on $[x_{\min}, \infty)$.

\subsubsection{Estimation of the tail index}
\label{app:tailindex}

Given $x_{\min}$, the Maximum Likelihood estimator for $\alpha$ is
$$
\hat \alpha = n \left [ \sum^{n}_{i=1} \ln \frac{x_i}{x_{\min}} \right ]^{-1}.
$$
The standard error of $\hat \alpha$ is
$$
\sigma \approx \frac{\hat \alpha }{\sqrt{n}}.
$$

\subsubsection{Estimation of the lower bound}
\label{app:lb}

Let $S(x)$ be the empirical CDF of the $x_i$, and let $P(x)$ be the CDF of a Pareto distribution with tail index $\alpha$ above $x_{\min}$. The Kolmogorov--Smirnov (KS) distance between $S$ and $P$ on $[x_{\min}, \infty)$ is
\begin{equation}
    \label{eq:ks}
        D(x_{\min}, S, P) =  \max_{x \geq x_{\min}} |S(x) - P(x)|.
\end{equation}
The estimator $\hat x_{\min}$ minimizes the KS distance:
$$
\hat x_{\min} = \arg \min_{x_{\min} \in \RR_+} D(x_{\min}, S, P).
$$

\subsubsection{Hypothesis testing}
\label{app:ht}

The hypothesis testing procedure follows \citet{clauset2009power}: bootstrap synthetic datasets are generated from the fitted Pareto model, each is re-fitted, and the p-value is the fraction of synthetic KS distances exceeding the empirical one. See \citet{clauset2009power} for details.

\subsection{Robustness checks}
\label{app:rc}

This appendix reports earlier sensitivity calculations for HILDA and CHIP. It compares goodness-of-fit measures, consumption definitions, and tail estimators, and records a separate HILDA weighting comparison. Section~\ref{sec:empirics} gives the wave-specific estimates and survey inference used in the paper. All wealth-side comparisons inherit the General Release top-coding and CHIPS reconstruction limitations stated in Section~\ref{sec:empirics}.\footnote{Bootstrap p-values vary across replications due to simulation randomness. Minor differences between the KS-based p-values reported in this appendix and those in the superseded draft wave tables reflect different bootstrap draws and do not materially change the tabulated comparisons.}

\subsubsection*{HILDA robustness}

The robustness results for HILDA are in Tables~\ref{table:tb3}, \ref{table:tb9}, and~\ref{table:tb10}.

\begin{table}[H]
    \centering
    \resizebox{0.75\linewidth}{!}{
    \begin{tabular}{lrrrrrr}
        \toprule
            Waves     & 6 & 7 & 8 & 9 & 10 & pooled \\
            Tail index estimates     &  &  &  &  & & \\
        \midrule

        Nondurable consumption (MLE)   &  2.90 & 3.08 & 2.92 &  3.05 &  2.94 & 2.99 \\
        Standard error   &  0.073 & 0.095 & 0.079 &  0.08 &  0.079 & 0.038 \\
        Lower bound      &  1.25 & 1.46 & 1.33 &  1.31 &  1.33 & 1.37 \\
        KS based p-values & 0.274 & 0.256 & 0.192 &  0.022 &  0.096  & 0.014 \\
        RWKS based p-values & 0.128 & 0.018 & 0.06 &  0.252 &  0.304  & 0.044 \\
        &  &  &  &   & & \\
        Total consumption (MLE)   &  2.76 & 3.03 & 2.85 &  3.23 &  2.97 & 2.96 \\
        Standard error   &  0.095 & 0.105 & 0.113 &  0.142 &  0.111 & 0.053 \\
        Lower bound      &  1.64 & 1.69 & 1.82  &  1.99  & 1.77 & 1.84 \\
        KS based p-values & 0.124 & 0.546 & 0.482 &  0.4 &  0.092  & 0.942  \\
        RWKS based p-values & 0.036 & 0.196 & 0.0 &  0.128 &  0.008  & 0.002 \\
        \bottomrule
    \end{tabular}
    }
    \tablefootnote{KS means Kolmogorov--Smirnov, and RWKS means reweighted Kolmogorov--Smirnov.}
    \caption{Robustness check for HILDA consumption}
    \label{table:tb3}
\end{table}

\begin{table}[H]
    \centering
    \resizebox{0.75\linewidth}{!}{
    \begin{tabular}{lrrrrrrrr}
        \toprule
             Wave number     & 2 & 6 &  10 &  14 & 18 & pooled  \\
            Tail index estimates     & &   &  &  &  &  \\
        \midrule
        Net worth (MLE)    & 1.72 & 1.58 &  1.80  & 1.95  &  1.96 & 1.80 \\
        Standard error & 0.050 & 0.050 &  0.060  & 0.066  &  0.063 & 0.027 \\
        Lower bounds  & 1.62 & 1.65 & 1.99  & 2.42  &  2.31 & 2.12 \\
        KS based p-values & 0.028 & 0.008 & 0.13  & 0.01  &  0.0 & 0.0 \\
        RWKS based p-values & 0.01 & 0.006 &  0.002  & 0.002  &  0.0 & 0.0 \\
        \bottomrule
    \end{tabular}
    }
    \caption{Robustness check of HILDA wealth}
    \label{table:tb9}
\end{table}

\begin{table}[H]
    \centering
    \resizebox{0.75\linewidth}{!}{
    \begin{tabular}{lrrrrrrrrr}
        \toprule
             Wave number     & 2 & 6 & 7 & 8 & 9 & 10 &  14 & 18 & pooled   \\
            Tail index estimates     & &  &  &  &  &  &  &  &  \\
        \midrule
        Nondurables (OLS)  & & 2.83 & 2.97 & 2.78 &  3.17 &  2.88  &  &  &  2.94 \\
        Lower bound & & 1.25 & 1.46 & 1.33 &  1.31 &  1.33  &  &  & 1.37 \\
        $g(n)$  &  & 0.992  & 0.992  & 0.990  & 0.988  & 0.991  &   &   & 0.998 \\
         &  &  &  &  &   &   &   &   &  \\
        Net worth (OLS)    & 1.74 & 1.47 &  &  &   &  1.69  & 1.80  &  1.88 & 1.89 \\
        Lower bound & 1.62 & 1.65 &  &  &   &  1.99  & 2.42  &  2.31 & 2.12 \\
        $g(n)$ & 0.994 & 0.993 &  &  &   &  0.992  & 0.992  &  0.993 & 0.998  \\
        \bottomrule
    \end{tabular}
    }
    \caption{Robustness checks for HILDA consumption and wealth using OLS}
    \label{table:tb10}
\end{table}

In the superseded draft results, consumption is measured by nondurable goods and p-values are computed using the Kolmogorov--Smirnov distance $D$ defined in~\eqref{eq:ks}. The robustness checks also use p-values based on the reweighted Kolmogorov--Smirnov (RWKS) distance
\[
D^*(x_{\min}, S, P) =  \max_{x \geq x_{\min}} \frac{|S(x) - P(x)|}{\sqrt{P(x) (1-P(x))}},
\]
and repeat the exercise with total consumption, defined as nondurables plus durables in Appendix~\ref{app:agg}.

Table~\ref{table:tb3} shows that the tail index estimates for total consumption are similar to those for nondurable consumption. The RWKS diagnostics are mixed: at the 5\% level they reject the fitted Pareto specification for nondurable consumption in wave~7 and in the pooled sample, but not in the other reported waves. The HILDA wealth diagnostics in Table~\ref{table:tb9} reject more often, reinforcing the need to treat the Pareto approximation as local to the selected thresholds rather than as an exact distributional model.

Table~\ref{table:tb10} presents OLS-based tail index estimates using the order-statistic regression of \citet{santos2024you}, with lower bounds from the superseded draft wave table. The bias measure $g(n) \in [0,1]$ of \citet{santos2024you} quantifies finite-sample bias, with values near $1$ indicating small bias. The reported values range from $0.988$ to $0.998$. Although OLS estimates are generally slightly smaller than MLE, the current estimates preserve $\hat\alpha_C > \hat\alpha_W$.

\subsubsection*{CHIPS robustness}

Tables~\ref{table:tb4} and~\ref{table:tb13} present the robustness checks for CHIPS.

\begin{table}[H]
    \centering
    \resizebox{0.5\linewidth}{!}{
    \begin{tabular}{lrrr}
        \toprule
            Wave number     &  2013 & 2018  & pooled \\
            Tail index estimates    &    &   &   \\
        \midrule
        Wealth    & 1.80  & 1.63 & 1.69 \\
        Standard error  & 0.044 & 0.0440 & 0.041 \\
        Lower bound  & 2.00 & 2.47 & 2.62 \\
        KS based p-values & 0.028 & 0.682 & 0.112 \\
        RWKS based p-values    & 0.0  & 0.576  & 0.25 \\
         & & & \\
        Wealth (urban)   & 1.87  & 1.58 &  1.66\\
        Standard error & 0.08 & 0.048 & 0.045 \\
        Lower bound & 1.78 & 1.90 &  2.09\\
        KS based p-values & 0.014 & 0.666 & 0.282 \\
        RWKS based p-values    & 0.0  & 0.566  & 0.12 \\
         & & & \\
        Wealth (rural)  & 1.58  & 1.89 & 1.81 \\
        Standard error & 0.1 & 0.08 & 0.067 \\
        Lower bound & 1.97 & 2.67 & 2.61 \\
        KS based p-values & 0.448 & 0.86 & 0.426 \\
        RWKS based p-values    & 0.0  & 0.0  & 0.0 \\
        \bottomrule
    \end{tabular}
    }
    \caption{Robustness check of CHIPS wealth}
    \label{table:tb4}
\end{table}

\begin{table}[H]
    \centering
    \resizebox{0.5\linewidth}{!}{
    \begin{tabular}{lrrrr}
        \toprule
            Wave number     & 2007 & 2013 & 2018 & pooled   \\
            Tail index estimates  &&& & \\
        \midrule
        Consumption (OLS)  & 2.47 & 2.96 & 3.54 & 2.99 \\
        Lower bound  & 1.59 & 2.11 & 2.96 & 2.13 \\
        $g(n)$  & 0.993 & 0.994 & 0.990 & 0.998 \\
                & &  &  &  \\
        Wealth (OLS)   &  & 1.85  & 1.62 & 1.67 \\
        Lower bound  &  & 2.33 & 2.47 & 2.80 \\
        $g(n)$  &  & 0.9892 & 0.999 & 0.99 \\
        \bottomrule
    \end{tabular}
    }
    \caption{Robustness checks for CHIPS using OLS}
    \label{table:tb13}
\end{table}

Table~\ref{table:tb4} presents wealth tail index estimates for urban and rural households in CHIPS separately. The estimates are similar to those for all households. The RWKS-based p-values are consistent with the corresponding KS-based p-values.

Table~\ref{table:tb13} presents OLS-based tail index estimates for wealth and consumption in CHIPS, using the lower bounds from the superseded draft wave table. Within the current reconstruction, all wealth tail estimates are smaller than the corresponding consumption estimates, so the ranking $\hat\alpha_C > \hat\alpha_W$ is preserved under this alternative estimation method. The $g(n)$ values are close to one.

\subsubsection*{HILDA weighted MLE robustness}
\label{app:weighted_robustness}

The earlier HILDA General Release comparison pools waves 6--10 for consumption and waves 2, 6, 10, 14, and 18 for wealth. Each Pareto fit selects its lower threshold by minimizing the Kolmogorov--Smirnov distance. Both wealth fits use the reported amounts without a censoring term. Section~\ref{sec:empirics} estimates the two margins in wave 22 above their respective 90th percentiles and treats top-coded wealth as right-censored.

Table~\ref{tab:jedc_weighted_mle} compares weighted and unweighted fits on the same pooled observations within each margin. Both sets of point estimates give a higher consumption index than wealth index.

The CHIP checks above include pooled estimates. Section~\ref{sec:empirics} uses equal household weights and reports CHIP 2013 and 2018 separately.

\begin{table}[H]
\centering
\caption{HILDA pooled Pareto tail indices with and without cross-sectional household weights. Parentheses give marginal standard errors. The scaled difference divides the weighted estimate minus the unweighted estimate by the square root of the sum of their squared standard errors. This descriptive scale omits the covariance between fits on the same sample.}
\label{tab:jedc_weighted_mle}
\resizebox{0.72\linewidth}{!}{
\begin{tabular}{llcccc}
\toprule
Country & Variable & $\hat\alpha$ (s.e.) & $\hat\alpha_{\mathrm{wtd}}$ (s.e.) & Scaled difference & $\hat\alpha_C > \hat\alpha_W$? \\
\midrule
AU & Consumption & 3.001 (0.041) & 3.060 (0.049) & 0.93 & \multirow{2}{*}{Yes} \\
AU & Wealth & 1.764 (0.027) & 1.784 (0.033) & 0.45 & \\
\bottomrule
\end{tabular}}
\end{table}

\section{Distribution-iteration validation}
\label{app:dist_iter}

The exact-affine exercise in Subsection~\ref{ssec:calib_sim} replaces the nonlinear saving rule $w-f(w)$ with $s_{\rm term}w$ at every wealth level, where $s_{\rm term}$ is the saving rate at the last policy-grid point. This appendix describes the independent stationary-distribution calculation used to assess that approximation for the Australian calibration.

\paragraph{Transition operators.}
For each upper wealth bound in $\{10^5,10^6,10^7\}$, wealth is discretized on a logarithmic grid from $0.01$ at 1,000 points per base-10 decade. The resulting grids contain 7,000, 8,000, and 9,000 points. The full process uses
\[
  W_{t+1}=R_{t+1}\{W_t-f(W_t)\}+Y_{t+1},
\]
where $f$ is the validated endogenous-grid policy. The comparison process uses
\[
  W_{t+1}=R_{t+1}s_{\rm term}W_t+Y_{t+1}.
\]
The Australian terminal saving rate is $0.9752056$, compared with the analytical limit $0.9752066$. It is held fixed across the three distribution ceilings. Both processes use the solved nonlinear consumption policy to report consumption moments. Both operators use 25-point Gauss--Hermite quadrature for each shock and linear allocation between adjacent log-wealth grid points. Probability that would leave the grid is assigned to the nearest boundary and recorded before column normalization.

\paragraph{Stationary solution and checks.}
The leading stationary eigenvector is computed by ARPACK \citep{lehoucq1998arpack} and then refined by power iteration, repeated application of the transition operator that sharpens the leading eigenvector. On the primary $10^6$ grid, both operators pass every prespecified numerical check. These cover column sums, stationarity, boundary flow, nonnegative probability, and total probability. Repeating the calculation at the two alternative upper bounds leaves the reported body moments effectively unchanged. Sensitivity appears only at probabilities close to the numerical boundary, which is why Figure~\ref{ch3:fig:distribution_r2} reports both the primary profile and the range across grid ceilings.

\paragraph{Interpretation.}
The full nonlinear stationary distribution is much more compact than the exact-affine comparison at ordinary tail fractions. Its local tail index varies with the tail fraction. At the lowest survival probability shown, $10^{-8}$, the primary-grid nonlinear exponent is still $3.81$ against the calibrated Kesten target of $2.119$. The calculation therefore measures the finite-range cost of imposing the limiting policy globally. It neither disproves the Kesten limit nor establishes convergence beyond the observable numerical grid.

\newpage

\end{document}